\documentclass[preprint,aps,amsmath,amssymb,showkeys,floatfix,longbibliography]{revtex4-2}

\newcommand{\cbox}[1]{%
  {\setlength{\fboxsep}{3pt}%
   \boxed{\vphantom{\tilde{G}(\omega)}\,#1\,}}}

\usepackage{graphicx}
\usepackage{hyperref}
\usepackage[capitalize]{cleveref}
\crefname{equation}{Eq.}{Eqs.}
\Crefname{equation}{Eq.}{Eqs.}

\begin{document}

\title{Closed-Form Analytical Charge Response Model for Silicon Photomultipliers
  with Recursive Correlated Avalanches}

\affiliation{School of Physical Sciences, University of Chinese Academy of Sciences, Beijing 101408, China}
\affiliation{Department of Engineering Physics, Tsinghua University, Beijing 100084, China}

\author{Yiqi Liu}
\author{Xuewei Liu}
\affiliation{Department of Engineering Physics, Tsinghua University, Beijing 100084, China}
\author{Benda Xu}
\email[Corresponding author, ]{orv@tsinghua.edu.cn}
\affiliation{School of Physical Sciences, University of Chinese Academy of Sciences, Beijing 101408, China}
\affiliation{Department of Engineering Physics, Tsinghua University, Beijing 100084, China}

\begin{abstract}
Silicon photomultipliers (SiPMs) have become the preferred photodetectors
in next-generation neutrino experiments, yet
no unified closed-form analytical expression free of truncation and numerical
convolution has been established for their full charge response spectrum,
which must simultaneously capture correlated cross-talk and afterpulsing
effects absent in conventional photomultiplier tubes (PMTs).
We present a unified closed-form model for the SiPM charge response within
the characteristic-function framework, treating pedestal noise,
single-electron-response (SER) charge, internal optical cross-talk, and afterpulsing on
equal footing.
The characteristic-function representation factorises the full charge
spectrum into three independent physical components: pedestal, single-electron
response (SER), and avalanche count statistics.
Prompt internal optical cross-talk is modelled as a Galton--Watson branching
process with Poisson offspring; building on the Generalised Poisson count
statistics identified by Vinogradov, we derive a Lambert $W$ closed form for
the total-progeny PGF via Lagrange--B\"{u}rmann inversion, providing the
analytical handle needed for efficient event-level reconstruction.
Afterpulsing is modelled as a per-avalanche geometric chain, derived as the
maximum-entropy Poisson--Gamma mixture: the exponential prior---maximum-entropy
for a positive continuous yield with fixed mean---marginalised over a Poisson
count yields the geometric per-avalanche distribution, whose $N$-avalanche
total is Negative Binomial. This naturally encompasses the Poisson afterpulsing
limit and recursive afterpulse chains while preserving analytical closure.
The resulting eight-parameter expression is further applied to
derive an explicit per-channel charge-time likelihood for event-level energy
reconstruction without numerical convolution at inference time.
\end{abstract}

\keywords{Silicon photomultiplier; charge response; Generalized Poisson;
  Lambert $W$ function; afterpulsing; Negative Binomial; event
  reconstruction}

\maketitle

\section{Introduction}

Silicon photomultipliers (SiPMs) are arrays of single-photon avalanche
diodes (SPADs) operated in Geiger mode above the breakdown voltage.
Compared with conventional photomultiplier tubes (PMTs), SiPMs offer
high photon detection efficiency, compact geometry, cryogenic
compatibility, and immunity to magnetic fields.
These features have made them the photosensor of choice
in neutrino and rare-event detectors including JUNO-TAO, DUNE, and
SBND, as well as neutrinoless-double-beta searches such as NEXT and
nEXO~\cite{venettacci_sipm_2024,falcone_cryogenic_2021,%
  sbnd_scintillation_2024,herrero_readout_2012,gallina_performance_2022}.

Accurate modelling of the \emph{charge response}, the probability
distribution of the integrated charge per event, underpins energy
calibration, trigger efficiency determination, and event-vertex
reconstruction in these detectors.
Because the charge response enters directly as the per-channel likelihood
in maximum-likelihood reconstruction, an analytical closed form is needed
for detectors with $\mathcal{O}(10^4)$ readout channels.
For PMTs, each photoelectron initiates a cascade multiplication whose
output charge converges to a Gamma single-electron response (SER) in the
high-gain limit~\cite{wright_low_1954,baldwin_statistics_1965,prescott_statistical_1966};
the Gamma distribution captures the asymmetric peak shape and
correct sub-PE valley density that Gaussian approximations miss.
The resulting compound-Poisson Gamma charge spectrum belongs to the
Tweedie exponential-dispersion family~\cite{jorgensen_theory_1997,dunn_generalized_2018},
which has been introduced in the 8-inch MCP-PMT modelling by Weng et al.~\cite{weng_compound_2024}

SiPMs exhibit three correlated effects absent in PMTs.
\emph{Internal optical cross-talk}: hot-carrier luminescence during an
avalanche emits secondary photons absorbed by neighbouring SPADs,
triggering discharges
recursively~\cite{duAfterpulsingCrosstalkMultipixel2008,wangMechanismsInternalCrosstalk2025}.
\emph{Afterpulsing}: carriers trapped at lattice defects are released and
trigger delayed avalanches~\cite{duAfterpulsingCrosstalkMultipixel2008,%
  jhaSimulatingSiliconPhotomultiplier2013,%
  butcherMethodCharacterizingAfterpulsing2017}.
\emph{Dark counts}: thermally generated carriers fire independent Geiger
discharges at random gate times, filling the $K\in(0,1)$ inter-peak
valley~\cite{chmill_characterisation_2017}.
Established SiPM charge spectrum models, including semi-analytical
convolution approaches~\cite{chmill_characterisation_2017} and the
four-building-block framework of
\textsc{PeakOTron}~\cite{rolph_peakotron_2023}, handle these
correlations through explicit sums over discrete PE or discharge counts,
truncated at a maximum $k_\text{max}$.
No closed-form treatment covering all three effects has been reported.

Characteristic-function methods applied to PMT calibration by Kalousis
et al.\ evaluate the full charge response as a single inverse DFT to all
orders in the Poisson mean, eliminating explicit convolution
integrals~\cite{kalousis_fast_2020,kalousis_calibration_2023}.
The analogous CF approach for SiPMs is non-trivial because cross-talk and
afterpulsing introduce inter-avalanche correlations absent in PMTs.
With Generalised Poisson PE count (parameters $\lambda,\theta$), Negative
Binomial afterpulsing ($\rho$, afterpulse charge density $f_\mathrm{ap}$
Sec.~\ref{sec:ap-charge}), and Poisson dark counts ($\mu_d$,
single-dark-pulse density $d_1$, Sec.~\ref{sec:darkcounts}), the
direct-sum spectrum is
\begin{equation}
\begin{aligned}
  G(q) &= \sum_{n,A,i}\, w_{n,A,i}\,
           \bigl[g_0 * f^{*n} * f_\mathrm{ap}^{*A} * d_1^{*i}\bigr](q), \\
  w_{n,A,i} &\equiv p_n^{\mathrm{GP}}\,
               \binom{A+n-1}{A}(1-\rho)^n\rho^A \\
             &\quad\times\frac{e^{-\mu_d}\mu_d^i}{i!},
\end{aligned}
  \label{eq:direct_sum}
\end{equation}
a triple sum over PE count $n\leq k_\text{max}$, afterpulse count
$A\leq A_\text{max}$, and dark-pulse count $i\leq i_\text{dark}^\text{max}$,
each $(n,A,i)$ term requiring a separately stored convolution;
the combinatorial weight $\binom{A+n-1}{A}\sim A^{n-1}/(n-1)!$ and
$k_\text{max}\sim 3\lambda$--$5\lambda$ must scale with the signal level.

In event reconstruction $\lambda$ varies across channels; without a
principled bound on $k_\text{max}$, the direct sum~\cref{eq:direct_sum}
must be re-evaluated for every channel at inference time.

In this work we derive a characteristic-function (CF) framework that
replaces every summation in~\cref{eq:direct_sum} (over prompt
avalanche count, afterpulse count, and dark-pulse count) with a
closed-form expression evaluable at a single frequency grid point,
free of truncation parameters and combinatorial bookkeeping.
The physical model follows established SiPM physics (detailed in
Sec.~\ref{sec:framework}): Gamma SER, exponential SPAD voltage
recovery, and asynchronous dark-pulse charge density.
The closed-form spectrum provides both a calibration tool and the
basis for an explicit per-channel charge-time likelihood for
event-level reconstruction.

Building on Vinogradov's Generalised Poisson count
statistics~\cite{vinogradov_analytical_2012}, we derive a
Lambert~$W$ closed form for the Galton--Watson total-progeny probability
generating function (PGF) via Lagrange--B\"{u}rmann
inversion~\cite{consul_generalized_1973,corless_lambert_1996}.
This closed form allows direct evaluation of the operator
$(1+\partial_\lambda)G$ in per-channel likelihood evaluation and analytic
$\lambda$-factorisation in calibration pre-computation
(Sec.~\ref{sec:reconstruction}).

Afterpulsing is treated with an explicit count-and-charge model rather than a
single-parameter approximation.
The per-avalanche afterpulse count follows a Geometric chain derived as
the maximum-entropy Poisson--Gamma mixture with an exponential prior on
per-microcell yield; summed over $N$ avalanches the total count is
Negative Binomial.
The afterpulse charge fraction follows $\mathrm{Beta}(2,b)$, derived
from the SPAD voltage-recovery curve (Sec.~\ref{sec:afterpulsing}).

Dark counts enter as an independent compound-Poisson factor
$\tilde{D}(\omega)$, replacing truncated auto-convolution series
(Sec.~\ref{sec:darkcounts}).

The closed-form expression yields an explicit
per-channel charge-time likelihood for event-level energy
reconstruction without numerical convolution at inference time
(Sec.~\ref{sec:reconstruction}).

\section{Charge Response Framework}\label{sec:framework}

\subsection{Convolution structure}

The measured charge in a readout window receives three statistically independent
contributions: electronic baseline noise (pedestal), signal-induced avalanche charge
from photoelectron-triggered SPADs, and thermally generated dark pulses.
Since independent charge contributions add, the full charge-response probability
density is their convolution:
\begin{equation}
  G_{\mathrm{full}}(q) = \Bigl(
    \underset{\text{noise floor}}{\cbox{g_0}}
    *
    \underset{\text{signal-induced}}{\cbox{G_{\mathrm{sig}}}}
    *
    \underset{\text{thermally induced}}{\cbox{G_{\mathrm{dark}}}}
  \Bigr)(q).
  \label{eq:G_full_q}
\end{equation}
The \emph{noise floor} $g_0(q)$ is the Gaussian pedestal from electronic white
noise accumulated over the readout window:
\begin{equation}
  g_0(q) = \frac{1}{\sqrt{2\pi}\,\sigma_0}
  \exp\!\left(-\frac{(q - m_0)^2}{2\sigma_0^2}\right),
  \label{eq:g0}
\end{equation}
with baseline offset $m_0$ and width $\sigma_0$.
The \emph{signal-induced} term $G_{\mathrm{sig}}(q)$ assembles charges from $N$
photoelectron-triggered avalanche clusters; writing $f_{\mathrm{eff}}(q)$ for the
effective single-cluster charge distribution (including afterpulses,
Sec.~\ref{sec:afterpulsing}),
\begin{equation}
\begin{aligned}
  G_{\mathrm{sig}}(q) &= \sum_{k=0}^{\infty} p_k\, f_{\mathrm{eff}}^{*k}(q), \\
  \quad p_k &= \mathbb{P}(N=k),
\end{aligned}
  \label{eq:Gsig}
\end{equation}
where $f_{\mathrm{eff}}^{*k}$ is the $k$-fold self-convolution
($f_{\mathrm{eff}}^{*0} = \delta$).
The \emph{thermally induced} term $G_{\mathrm{dark}}(q)$ is the charge distribution
of independent Poisson-distributed dark pulses, derived in Sec.~\ref{sec:darkcounts}.

\textsc{PeakOTron}~\cite{rolph_peakotron_2023} implements
\cref{eq:G_full_q}--\cref{eq:Gsig} with four physical building blocks:
\begin{itemize}
  \item[(i)] Gaussian SER with per-peak width
    $\sigma_k=\sqrt{\sigma_0^2+k\sigma_1^2}$;
  \item[(ii)] Binomial afterpulse count
    $J\!\mid\!k\sim\mathrm{Binomial}(k,p_\mathrm{ap})$
    per $k$ avalanches, with charge drawn from the SPAD
    exponential-recovery curve;
  \item[(iii)] dark-count charge assembled from charge-domain
    auto-convolutions of the single-dark-pulse density, truncated at
    $i_\mathrm{dark}^\mathrm{max}$ terms; and
  \item[(iv)] direct-sum evaluation of \cref{eq:Gsig} truncated at
    $i_\gamma^\mathrm{max}$ photon-induced PE terms.
\end{itemize}
The truncation in (iii)--(iv) and the separately stored convolutions in (i)
are resolved by the characteristic-function (CF) framework. The convolution
theorem converts \cref{eq:G_full_q} into three independent multiplicative
factors, each computable at $\mathcal{O}(N_\omega)$ cost, free of truncation
parameters.

\subsection{Characteristic-function factorisation}

Since convolution in position space becomes pointwise multiplication in
frequency space, the three statistically independent contributions in
\cref{eq:G_full_q} each supply one multiplicative factor:
\begin{equation}
  \tilde{G}_{\mathrm{full}}(\omega)
  = \underset{\text{noise floor}}{\cbox{\tilde{g}_0(\omega)}}
  \cdot
  \underset{\text{signal-induced}}{\cbox{G_N\!\bigl(\tilde{f}_{\mathrm{eff}}(\omega)\bigr)}}
  \cdot
  \underset{\text{thermally induced}}{\cbox{\tilde{D}(\omega)}},
  \label{eq:G_full_preview}
\end{equation}
where $\tilde{g}_0$, $G_N(\tilde{f}_{\mathrm{eff}})$, and $\tilde{D}$ are the
CFs of $g_0$, $G_{\mathrm{sig}}$, and $G_{\mathrm{dark}}$ respectively.
The factorisation of the signal-induced CF follows from \cref{eq:Gsig}.
Since the CF of a convolution is the product of CFs,
$\widetilde{f_{\mathrm{eff}}^{*k}} = \tilde{f}_{\mathrm{eff}}^k$, and therefore
\begin{equation}
\begin{aligned}
  \widetilde{G_{\mathrm{sig}}}(\omega)
  &= \sum_{k=0}^{\infty} p_k\,\tilde{f}_{\mathrm{eff}}(\omega)^k \\
  &= G_N\!\bigl(\tilde{f}_{\mathrm{eff}}(\omega)\bigr),
\end{aligned}
  \label{eq:factorisation}
\end{equation}
where $G_N(s) = \mathbb{E}[s^N] = \sum_{k=0}^{\infty} p_k s^k$ is the
probability generating function (PGF) of $N$.
The three factors in \cref{eq:factorisation} are derived independently in
Secs.~\ref{sec:crosstalk}--\ref{sec:darkcounts} and combine by pointwise
multiplication on the frequency grid.

In practice $\tilde{G}_{\mathrm{full}}$ is evaluated on a uniform frequency grid
with $N_\omega$ points; the inverse DFT recovers $G_{\mathrm{full}}$ on a charge
grid of spacing $\Delta q$.
The dominant aliasing error comes from the Gaussian pedestal,
\begin{equation}
  \bigl|\tilde{g}_0(\omega_\mathrm{N})\bigr|
  = \exp\!\left(-\frac{\sigma_0^2\pi^2}{2\,\Delta q^2}\right),
  \label{eq:aliasing}
\end{equation}
at the Nyquist frequency $\omega_\mathrm{N} = \pi/\Delta q$.
\Cref{eq:aliasing} is exact; the aliasing error is negligible whenever
$\Delta q \ll \sigma_0$, a condition easily satisfied in practice
independently of $\lambda$ or any truncation choice.

\subsection{Single-electron response (SER) charge distribution}\label{sec:spe}

The framework of \cref{eq:factorisation} accommodates any SER
distribution $f(q)$.
A single photoelectron initiates a cascade multiplication process in which
each stage is governed by a branching distribution; Baldwin and
Friedman~\cite{baldwin_statistics_1965} and
Prescott~\cite{prescott_statistical_1966} showed that the output charge
distribution converges to a Gamma distribution in the high-gain limit,
\begin{equation}
  f(q) = \frac{q^{\alpha-1}\,e^{-q/\beta}}{\beta^\alpha\,\Gamma(\alpha)},
  \qquad q > 0,
  \label{eq:gamma_spe}
\end{equation}
with shape parameter $\alpha > 0$ and scale $\beta > 0$.
The CF of the Gamma SPE is
\begin{equation}
  \tilde{f}(\omega) = (1 + i\beta\omega)^{-\alpha}.
  \label{eq:gamma_cf}
\end{equation}
The Gaussian approximation $\tilde{f}(\omega) \approx e^{i\omega\alpha\beta
- \alpha\beta^2\omega^2/2}$ is recovered in the limit $\alpha \to \infty$,
$\beta \to 0$ with $\alpha\beta$ fixed.
We adopt the Gamma SER model \cref{eq:gamma_spe,eq:gamma_cf}
as the primary example throughout; all results in subsequent
sections hold for any analytic $\tilde{f}$.

\subsection{Baseline: Poisson count statistics}\label{sec:poisson}

Without cross-talk or afterpulsing ($\theta=\rho=0$), $f_{\mathrm{eff}}=f$ and
$N$ follows Poisson with mean $\lambda$:
\begin{equation}
\begin{aligned}
  p_k &= e^{-\lambda}\frac{\lambda^k}{k!}, \\
  G_N(s) &= e^{\lambda(s-1)}.
\end{aligned}
  \label{eq:poisson_pgf}
\end{equation}
Substituting \cref{eq:poisson_pgf} and \cref{eq:gamma_cf} into \cref{eq:factorisation},
\begin{equation}
  \widetilde{G_{\mathrm{sig}}}(\omega)
  = \exp\!\Bigl(\lambda\bigl((1+i\beta\omega)^{-\alpha} - 1\bigr)\Bigr).
  \label{eq:poisson_cf}
\end{equation}
This is the signal-induced CF for the compound-Poisson Gamma distribution,
a member of the Tweedie exponential-dispersion
family~\cite{jorgensen_theory_1997} with power index
$p=(\alpha+2)/(\alpha+1)\in(1,2)$~\cite{dunn_generalized_2018},
and is the baseline for adding SiPM-specific effects~\cite{tao_reconstruction_2025}.

\subsection{Structure of the full model}

The gain-normalised charge $K\equiv(q-m_0)/G^*$, where $G^*=\alpha\beta$ is
the mean single-avalanche charge (gain), is used throughout: integer-PE
peaks occur at $K=n$ and inter-peak valleys at $K\in(n,n+1)$.
The effective gain $G^*$ reported in Sec.~\ref{sec:validation}
is the MLE estimate of $\alpha\beta$.

The three factors of~\cref{eq:G_full_preview} are built up in layers
on the Poisson baseline~\cref{eq:poisson_cf}.
Internal optical cross-talk (Sec.~\ref{sec:crosstalk}) replaces $G_N$ in
\cref{eq:factorisation} with the Generalised Poisson PGF via a Lambert~$W_0$
closed form for the total-progeny branching process.
Afterpulsing (Sec.~\ref{sec:afterpulsing}) folds per-avalanche secondary
avalanches into $\tilde{f}_\mathrm{eff}$~\cref{eq:f_eff}, preserving the
factorised form.
Dark counts (Sec.~\ref{sec:darkcounts}) enter as the statistically
independent factor $\tilde{D}(\omega)$~\cref{eq:D_cf}, replacing the
auto-convolution series in \cref{eq:direct_sum} with a closed-form expression
at $\mathcal{O}(N_\omega)$ cost.

\subsection{Scope and model limitations}

Several second-order effects lie outside this model.
External cross-talk, driven by photons reflected at the SiPM package surface
or generated in adjacent pixels, depends on the detector geometry and
housing response. It enters the detector-level map from
$(E,\,\vec{r})$ to the mean PE array $\{\lambda_j\}$; the present work
addresses only the mapping from $\{\lambda_j\}$ to the per-channel charge
likelihood.
Dark-induced correlated avalanches, in which a thermally generated discharge
triggers cross-talk secondaries or afterpulse chains, are neglected.
At the validated operating point (60\,V), $\mu_d = 0.103$ and
$\theta = 0.047$, giving a cascade probability
$\mu_d\theta \approx 0.005$ per gate.
The pull plots (Fig.~\ref{fig:chisq})
show no systematic bias from this higher-order contribution.
The model does not capture finite microcell saturation, which truncates
the Galton--Watson tree when neighbouring cells are already in avalanche.
This effect becomes appreciable only at high occupancy
($\lambda \mathbb{E}[N/\lambda] \gg N_\text{cell}$), well outside the
calibration regime.
Temperature dependence of DCR, cross-talk, and afterpulsing is not modelled
and must be re-calibrated at each operating temperature; this is relevant
for cryogenic deployments.
Multi-trap-species structure, which can produce non-exponential afterpulse
timing, is absorbed into an effective single-$\rho$ parameterisation.

\section{Internal Optical Cross-talk: Generalized Poisson}
\label{sec:crosstalk}

This section derives the avalanche count PGF $G_N(s)$ that drives the
signal-induced factor of \cref{eq:G_full_preview}.

\subsection{Physical mechanism and branching model}

During each SPAD avalanche, a fraction of the $\sim 10^8$--$10^9$ charge
carriers lose energy through hot-carrier luminescence, emitting secondary
photons in the near-infrared.
These photons may be absorbed by adjacent SPADs and trigger secondary
avalanches; each secondary avalanche can in turn emit photons, triggering
tertiary avalanches, and so on~\cite{duAfterpulsingCrosstalkMultipixel2008,wangMechanismsInternalCrosstalk2025}.
Among the several cross-talk mechanisms identified
in~\cite{wangMechanismsInternalCrosstalk2025}, the present model covers
\emph{internal} optical cross-talk in both its prompt form (photon
absorbed directly in a neighbouring SPAD's active region) and its delayed
form (photon absorbed in the neutral region, with the resulting carrier
diffusing into the depletion layer after a short delay).
In a charge-integration readout with a window much longer than the carrier
diffusion time, both mechanisms contribute a full Geiger discharge charge
and are indistinguishable; the effective cross-talk parameter
$\theta=\theta_{\mathrm{prompt}}+\theta_{\mathrm{delayed}}$ absorbs both.
External cross-talk driven by photons reflected at the SiPM package surface
or generated outside the active silicon layer is determined by the detector
geometry and housing response; it is typically modelled jointly with
surface-reflectance corrections and is outside the scope of this paper.
We model the internal optical cascade as a Galton--Watson branching tree: each
triggered SPAD independently emits $k$ secondary photons absorbed by
neighbouring cells, with $k\sim\mathrm{Poisson}(\theta)$, $\theta\in[0,1)$
parametrising the cross-talk strength ($\theta\approx0.004$--$0.047$
for the PCB6 device across 53--60\,V), and each newly triggered SPAD
initiates the same process.
For $\lambda$ primary photoelectrons, $\lambda$ independent such trees are
initiated; the total avalanche count $N$ is the combined progeny of all
trees.
The Galton--Watson model assumes unlimited cell availability; finite microcell
occupancy limits the total progeny to the number of unoccupied cells and
is not captured here.
This saturation correction becomes appreciable only when the mean total
avalanche count approaches the total microcell number $N_\text{cell}$,
which is well outside the calibration regime at the LED illumination
levels used in this work.

\subsection{Total-progeny PGF: fixed-point equation and Lambert $W$ closed form}

Let $T(s)$ denote the PGF of the total progeny of a single tree, including
the root.
Since the subtrees rooted at each of the root's offspring are independent
copies of the full tree, the PGF of the total progeny satisfies the standard
Galton--Watson functional equation~\cite{athreya_branching_1972}.
With the offspring PGF $\phi(s)=e^{\theta(s-1)}$ of $\mathrm{Poisson}(\theta)$
this gives
\begin{equation}
  T(s) = s\,\phi\bigl(T(s)\bigr) = s\,e^{\theta(T(s)-1)}.
  \label{eq:T_implicit}
\end{equation}
For $\theta<1$ the branching is subcritical and the tree terminates almost
surely; setting $s=0$ gives $T(0)=0$.
For $\lambda$ Poisson-distributed primary PE events, the combined
total-progeny PGF is the Poisson PGF evaluated at $T(s)$:
\begin{equation}
  G_N(s) = e^{\lambda(T(s)-1)}.
  \label{eq:gp_pgf_form}
\end{equation}
Setting $\theta=0$ gives $T(s)=s$, recovering the Poisson baseline
$G_N(s)=e^{\lambda(s-1)}$.
The PMF follows by Lagrange--B\"{u}rmann inversion of~\cref{eq:gp_pgf_form}.
Here $[s^k]\,(\cdot)$ denotes the coefficient of $s^k$ in the formal power
series of $(\cdot)$, and the inversion formula holds for any $H$ analytic
at $t=0$.
Setting $H(t)=e^{\lambda(t-1)}$ and $\phi(t)=e^{\theta(t-1)}$ in
$[s^k]H(T(s))=\frac{1}{k}[t^{k-1}](H'(t)\phi(t)^k)$,
\begin{equation}
  p_k = \frac{\lambda(\lambda+\theta k)^{k-1}\,e^{-\lambda-\theta k}}{k!},
  \quad k \geq 1,
  \label{eq:gp_pmf}
\end{equation}
the Generalised Poisson (GP)
distribution~\cite{consul_generalized_1973,vinogradov_analytical_2012},
with moments $\mathbb{E}[N]=\lambda/(1{-}\theta)$ and
$\mathrm{Var}(N)=\lambda/(1{-}\theta)^3$.
The GP distribution arises naturally from the Galton--Watson branching
structure: each primary photoelectron seeds an independent avalanche
tree, and summing over $\lambda$ independent trees (the Poisson-distributed
primaries) produces an over-dispersed count in which cross-talk inflates
both the mean and variance by the factor $(1-\theta)^{-1}$ and
$(1-\theta)^{-3}$, respectively.
The present work contributes the closed-form $T(s)$~\cref{eq:T_lambertW}.
\Cref{eq:T_implicit} can be solved in closed form.
Rewriting as $T\,e^{-\theta T}=s\,e^{-\theta}$ and multiplying both sides
by $-\theta$: $(-\theta T)\,e^{-\theta T}=-\theta s\,e^{-\theta}$.
By the definition $W_0(z\,e^z)=z$ of the principal branch $W_0$ of the
Lambert $W$ function~\cite{corless_lambert_1996}, real-valued for
arguments in $[-e^{-1},\infty)$,
\begin{equation}
  T(s) = -\frac{1}{\theta}\,W_0\!\bigl(-\theta s\,e^{-\theta}\bigr).
  \label{eq:T_lambertW}
\end{equation}
The boundary conditions $T(0)=0$ (since $W_0(0)=0$) and $T(1)=1$
(since $W_0(-\theta e^{-\theta})=-\theta$ by $W_0(xe^x)=x$ with $x=-\theta$)
are immediately verified.
As $\theta\to 1^-$, the argument $-\theta e^{-\theta}\to -e^{-1}$
approaches the branch point of $W_0$; the GP variance
$\mathrm{Var}(N)=\lambda/(1-\theta)^3$ diverges and the distribution
becomes heavy-tailed.
For real $\theta$ in the physical range $[0, 1)$, the argument
$-\theta e^{-\theta}$ is real and in $[-e^{-1}, 0]$, so $W_0$ takes
values in $[-1, 0]$ and is evaluated on the principal real branch.
In practice, $\theta \leq 0.20$ for standard SiPMs (Table~\ref{tab:params}),
so the argument stays far from the branch point.
Substituting \cref{eq:T_lambertW} into \cref{eq:gp_pgf_form} and then into \cref{eq:factorisation},
\begin{equation}
  \widetilde{G_{\mathrm{sig}}}(\omega)
  = \exp\!\left[\lambda\!\left(
  -\frac{W_0\!\bigl(-\theta\,\tilde{f}(\omega)\,e^{-\theta}\bigr)}{\theta}
  - 1\right)\right].
  \label{eq:crosstalk_cf}
\end{equation}
This is the closed-form signal-induced characteristic function for a SiPM
with internal optical cross-talk.

\section{Afterpulsing: Negative Binomial Extension}\label{sec:afterpulsing}

This section derives $\tilde{f}_\mathrm{eff}(\omega)$, the effective
single-avalanche characteristic function that replaces $\tilde{f}$ in
the signal-induced factor of \cref{eq:G_full_preview}.

\subsection{Per-avalanche geometric model}\label{sec:afterpulsing-geom}

Afterpulsing in SiPMs arises from charge carriers temporarily trapped at
crystal lattice defects during an avalanche and subsequently released,
triggering a delayed avalanche~\cite{duAfterpulsingCrosstalkMultipixel2008,
jhaSimulatingSiliconPhotomultiplier2013,
butcherMethodCharacterizingAfterpulsing2017}.
We model the afterpulse chain from each triggered avalanche, whether
initiated by a primary photoelectron or by optical cross-talk, as a
geometric process, following the physical picture of
\textsc{PeakOTron}~\cite{rolph_peakotron_2023} while extending it to
recursive chains.
In \textsc{PeakOTron} the per-trigger afterpulse count is
Binomial~\cite{rolph_peakotron_2023,duAfterpulsingCrosstalkMultipixel2008}---at
most one afterpulse per primary discharge---whereas the geometric model here
accounts for afterpulse-of-afterpulse recursion: the first afterpulse occurs with
probability $\rho$; given that, the second occurs with the same probability $\rho$;
and so on.
The total chain length from a single SPAD (the number of afterpulse
avalanches in that microcell, including those triggered recursively) follows
a geometric distribution,
\begin{equation}
\begin{aligned}
  &A_j \sim \mathrm{Geom}_0(\rho): \\
  &\mathbb{P}(A_j = a) = (1-\rho)\,\rho^a, \\
  &\qquad a = 0, 1, 2, \ldots,
\end{aligned}
  \label{eq:geom}
\end{equation}
with $\rho \in [0,1)$.
The geometric distribution is the unique discrete memoryless
distribution. The probability $\rho$ of triggering the next afterpulse
does not depend on how many have already fired, because carrier
release from a trap is a Poisson process. Each carrier waits an
exponentially distributed time before release, and the resulting count
is geometric when sampled against a fixed readout window.
Real devices host multiple trap species with distinct release-time
constants~\cite{duAfterpulsingCrosstalkMultipixel2008}; the
single-$\rho$ parameterisation aggregates them into an effective rate,
absorbing species heterogeneity into a broader geometric distribution.
Setting $\rho = 0$ recovers the afterpulse-free limit.
In the limit $\rho\to 0$ the chain terminates almost surely after at most
one step: $\mathrm{Geom}_0(\rho)$ reduces to $\mathrm{Bernoulli}(\rho)$,
and the total afterpulse count summed over $N$ avalanches reduces to
$\mathrm{Binomial}(N,\rho)$---the explicit model of
\textsc{PeakOTron}~\cite{rolph_peakotron_2023} and the implicit assumption
in analyses reporting only a single afterpulse probability per
trigger~\cite{butcherMethodCharacterizingAfterpulsing2017,
duAfterpulsingCrosstalkMultipixel2008}.
Wang~\cite{hanwen_mass_testing} independently applies the geometric model
to afterpulse chains when characterising TAO SiPMs at scale.

\subsection{Negative Binomial total afterpulse count}

Let $N$ denote the total prompt avalanche count (distributed as Generalised
Poisson with parameters $\lambda,\theta$ from Sec.~\ref{sec:crosstalk}).
Conditional on $N = n$, the afterpulse-chain lengths $A_1, \ldots, A_n$ from
each prompt avalanche---including those triggered by cross-talk---are
i.i.d.~$\mathrm{Geom}_0(\rho)$, so conditioning on $N = n$ reduces the problem to summing $n$ independent
geometric random variables.
Their sum $M_n = \sum_{j=1}^n A_j$ has PGF equal to the $n$-fold
product,
\begin{equation}
  G_{M_n | N=n}(s)
  = \left(\frac{1-\rho}{1-\rho s}\right)^n,
  \label{eq:nb_pgf_cond}
\end{equation}
identifying $M_n | N=n$ as Negative Binomial with parameters $(n, 1-\rho)$.
The PMF is
\begin{equation}
\begin{aligned}
  \mathbb{P}(M_n = m \mid N = n)
  &= \binom{m+n-1}{m}(1-\rho)^n\rho^m, \\
  &\quad m = 0,1,2,\ldots,
\end{aligned}
  \label{eq:nb_pmf}
\end{equation}
with mean $n\rho/(1-\rho)$ and variance $n\rho/(1-\rho)^2$;
setting $n=1$ recovers \cref{eq:geom}.

Wang~\cite{hanwen_mass_testing}, characterising TAO SiPMs at scale,
treats the NPE avalanche cascade as a \emph{bundle} and applies a
single geometric process to the event as a whole.
The present model instead resolves the cascade into $N$ independent
i.i.d.\ Geom$_0(\rho)$ chains---one per triggered avalanche---thereby
accounting for afterpulse-count discrepancies across individual
avalanches; the Negative Binomial \cref{eq:nb_pgf_cond} is the
consequence of this per-avalanche decomposition.
Wang's model also omits the dark-count contribution $\tilde{D}(\omega)$
(Sec.~\ref{sec:darkcounts}), which
\textsc{PeakOTron}~\cite{rolph_peakotron_2023} includes explicitly; the
absence of $\tilde{D}$ leaves the continuous charge density in the
$K\in(0,1)$ valley unmodelled, leading to systematic underfitting of that
region and a compensating bias in the afterpulse parameters.
The count and charge degrees of freedom are treated separately:
Sec.~\ref{sec:ap-charge} derives the afterpulse charge distribution
from the SPAD voltage-recovery curve, following the physical picture
of \textsc{PeakOTron}~\cite{rolph_peakotron_2023}.

\subsection{Maximum-entropy interpretation}\label{sec:maxentropy}

The per-trigger afterpulse count is often modelled as Poisson-distributed
with a voltage-dependent
mean~\cite{duAfterpulsingCrosstalkMultipixel2008,chmill_characterisation_2017}.
Both models constrain the same physical mean $\mu=\rho/(1-\rho)$, set by
trap density and overvoltage at constant bias voltage.
They differ in the per-cell Poisson rate $\lambda_j$: because carrier
release is a Poisson process with exponentially distributed inter-release
times (Sec.~\ref{sec:afterpulsing-geom}), the maximum-entropy prior on the
per-cell rate with fixed mean $\mu$ is $\lambda_j\sim\mathrm{Exp}(\mu)$.
Marginalising $\mathrm{Poisson}(\lambda_j)$ over $\mathrm{Exp}(\mu)$,
\begin{equation}
\begin{aligned}
  \mathbb{P}(A_j=a)
  &= \int_0^\infty \frac{\lambda^a e^{-\lambda}}{a!}
     \cdot\frac{e^{-\lambda/\mu}}{\mu}\,\mathrm{d}\lambda \\
  &= (1-\rho)\,\rho^a, \quad \rho = \frac{\mu}{1+\mu},
\end{aligned}
  \label{eq:geom_marginalisation}
\end{equation}
recovering $\mathrm{Geom}_0(\rho)$~\cref{eq:geom}.
This \emph{exponential dispersion} of the per-cell Poisson yield bridges the
Negative Binomial and Poisson families:
$\mathrm{NegBin}(1,1{-}\rho)=\mathrm{Geom}_0(\rho)$ at $r=1$, while
$\mathrm{NegBin}(r,1{-}\rho)\to\mathrm{Poisson}(\mu)$ as $r\to\infty$ with
mean $\mu$ fixed---the Poisson count is the many-trap limit of the same
family, and the geometric the maximum-entropy choice when trap microstructure
is not resolved.
The geometric model is therefore a parsimonious closure that introduces
no trap-structure parameters beyond the mean $\mu$.
It preserves analytical tractability and encompasses both the
Poisson limit (many traps, $r\to\infty$) and the single-trap limit
($r=1$, Bernoulli in the low-$\rho$ regime).
At $\rho \lesssim 0.08$ as found here (Table~\ref{tab:params}), all
three models give nearly identical distributions, so the afterpulsing
model choice has little effect on fitted spectra.
The distinction becomes material only at $\rho \gtrsim 0.3$ or when
consecutive afterpulse chains are individually resolved.

\subsection{Afterpulse charge distribution}\label{sec:ap-charge}

The counting and charge degrees of freedom are treated separately: given
the Negative Binomial total afterpulse count from
Sec.~\ref{sec:afterpulsing}, each afterpulse $i$ contributes an
independent charge $Q_{\mathrm{ap},i}=G^* X_i$, where $G^*$ is the mean
single-avalanche charge (gain) and $X_i\in(0,1)$ is the fraction of the
full avalanche charge available at the moment of firing.
After a primary SPAD discharge at $t=0$, the SPAD recharges as
$x(t)=1-e^{-t/\tau_r}$, where $\tau_r$ is the voltage-recovery time
constant.
A trapped carrier released at $T_0\sim\mathrm{Exp}(1/\tau_\mathrm{Ap})$
fires a new discharge with probability governed by the \emph{trigger
efficiency} $\eta\!\bigl(x(T_0)\bigr)$, where $\eta:(0,1)\to(0,1)$
encodes how Geiger-discharge probability depends on recovered overvoltage.
Changing variables to $X=x(T_0)$ and size-biasing by $\eta$,
\begin{equation}
  f_X(x) \propto \eta(x)\,(1-x)^{b-1},
  \qquad b=\tau_r/\tau_\mathrm{Ap}.
  \label{eq:ap_charge_pdf}
\end{equation}
Multiple physically motivated forms of $\eta$ are consistent with available
data.
The minimal choice $\eta(x)=x$---trigger probability proportional to
recovered overvoltage~\cite{rolph_peakotron_2023}---reduces the charge
fraction to a single parameter $b = \tau_r/\tau_\mathrm{Ap}$:
\begin{equation}
  X_i \sim \mathrm{Beta}(2,\,b).
  \label{eq:beta_ap}
\end{equation}
The parameter $b = \tau_r/\tau_\mathrm{Ap}$ is the ratio of the SPAD
voltage-recovery timescale to the trap-release timescale.
Larger $b$ (fast trap release relative to recovery) concentrates afterpulses
at small $x$, reducing the mean charge fraction
$\langle X\rangle = 2/(2+b)$; smaller $b$ (slow release) allows the SPAD to
recover further before firing, increasing $\langle X\rangle$ toward 1.
The two-parameter choice $\eta(x)=[1-(1-x)^a]$ with
$a=\tau_r/\tau_\mathrm{rec}$~\cite{rolph_peakotron_2023} yields
\begin{equation}
  f_X(x) \propto
  \bigl[1-(1-x)^a\bigr]\,(1-x)^{b-1},
  \label{eq:ap_charge_general}
\end{equation}
where $a$ encodes the voltage-dependence of the Geiger-breakdown probability;
$\mathrm{Beta}(2,b)$ is the $a=1$ case.
We adopt $a=1$ throughout, yielding a single-parameter charge model
whose BIC performance matches that of the two-parameter form
within the surveyed voltage range.
When a systematic residual charge mismatch persists in the inter-PE valleys,
the two-parameter form \cref{eq:ap_charge_general} provides an additional
degree of freedom; BIC can then determine whether the extension is
warranted.
The fixed-charge limit $Q_\mathrm{ap}=G^*$ is recovered as $b\to 0$.
The afterpulse charge characteristic function is
\begin{align}
  \phi_\mathrm{ap}(\omega)
  &= \mathbb{E}_{X\sim\mathrm{Beta}(2,b)}\!\bigl[e^{i\omega G^* X}\bigr]
  \notag\\
  &= {}_1F_1\!\bigl(2;\,b+2;\,i\omega G^*\bigr),
  \label{eq:phi_ap}
\end{align}
the Kummer confluent hypergeometric function.

\subsection{Full closed-form charge spectrum}

The charge model combines the two-stage cascade: the prompt avalanche
count $N$ follows the Generalised Poisson distribution from cross-talk
(\S\ref{sec:crosstalk}), afterpulsing is then layered independently
per avalanche to yield the Negative Binomial total count
(\S\ref{sec:afterpulsing}), and each afterpulse charge is an independent
$\mathrm{Beta}(2,b)$ fraction of the gain (\S\ref{sec:ap-charge}).
Each prompt avalanche $j$ contributes a cluster of charges: the prompt
charge drawn from the SPE distribution $f(q)$ plus $A_j$ afterpulse
charges each independently drawn from $G^*\,\mathrm{Beta}(2,b)$.
The geometric series for the per-avalanche afterpulse contribution
sums to
\begin{align}
  \tilde{c}(\omega)
  &= \tilde{f}(\omega)\,\sum_{a=0}^{\infty}(1-\rho)\rho^a\,
    \phi_\mathrm{ap}(\omega)^a
  \notag\\
  &= \tilde{f}(\omega)\cdot
    \frac{1-\rho}{1-\rho\,\phi_\mathrm{ap}(\omega)}.
  \label{eq:c_eff}
\end{align}
Defining the effective SPE characteristic function
\begin{equation}
  \tilde{f}_{\mathrm{eff}}(\omega)
  := \tilde{f}(\omega)\cdot
  \frac{1-\rho}{1-\rho\,\phi_\mathrm{ap}(\omega)},
  \label{eq:f_eff}
\end{equation}
the total charge spectrum's signal-induced CF retains the factorised form \cref{eq:factorisation} with $\tilde{f}$ replaced by $\tilde{f}_\text{eff}$:
\begin{equation}
  \widetilde{G_{\mathrm{sig}}}(\omega)
  = G_N\!\bigl(\tilde{f}_{\mathrm{eff}}(\omega)\bigr).
  \label{eq:G_full}
\end{equation}
Substituting the Generalised Poisson PGF \cref{eq:gp_pgf_form} with
\cref{eq:T_lambertW} into \cref{eq:G_full} yields the signal-induced closed-form
characteristic function:
\begin{equation}
  \widetilde{G_{\mathrm{sig}}}(\omega)
  = \exp\!\left[\lambda\!\left(
    -\frac{W_0\!\!\left(-\theta\,e^{-\theta}\,
      \tilde{f}_{\mathrm{eff}}(\omega)\right)}{\theta} - 1
  \right)\right].
  \label{eq:G_full_explicit}
\end{equation}
This eight-parameter expression ($m_0$, $\sigma_0$, $\alpha$, $\beta$,
$\lambda$, $\theta$, $\rho$, $b$) encompasses all previously discussed
special cases:
\begin{itemize}
  \item $\theta = 0$, $\rho = 0$: Poisson + Gamma, recovering
    \cref{eq:poisson_cf};
  \item $\theta > 0$, $\rho = 0$: cross-talk only,
    recovering \cref{eq:crosstalk_cf};
  \item $\theta = 0$, $\rho > 0$: Poisson count with NB afterpulsing;
  \item $\theta > 0$, $\rho > 0$: full model \cref{eq:G_full_explicit}.
\end{itemize}
With Gamma SER \cref{eq:gamma_cf}, \cref{eq:G_full_explicit} becomes
fully explicit; with Gaussian SER one substitutes
$\tilde{f}(\omega) = e^{i\omega\mu_1 - \sigma_1^2\omega^2/2}$.

\section{Dark-Count Contribution}\label{sec:darkcounts}

This section derives the thermally induced factor $\tilde{D}(\omega)$ in
\cref{eq:G_full_preview}.

\subsection{Physical model}

Dark counts arise from thermally generated or field-assisted electron--hole
pairs that trigger Geiger discharges independently of the incident light
pulse.
In contrast to light-induced discharges, which occur synchronously with the
light pulse and contribute charge at integer multiples of the gain~$G^*$,
dark pulses arrive at random times relative to the integration gate
$[0,T]$.
A dark pulse at time $s$ contributes only the fraction of its exponential
charge transient that falls within the gate, yielding a sub-PE charge
fraction $a(s)\in(0,1)$.
This asynchronous partial integration is the origin of the continuous charge
density between the pedestal and 1PE peaks---the 0PE--1PE valley---observed
in SiPM spectra at elevated dark-count rates~\cite{chmill_characterisation_2017,
rolph_peakotron_2023}.

\subsection{Single-dark-pulse charge distribution and valley density}
\label{sec:dark-pdf}

For an exponential pulse shape with slow time constant $\tau$, a dark
discharge at time $s \in (-t_0,\,T)$ contributes integrated charge
$G^*\,a(s)$ to the gate, where
\begin{equation}
  a(s) =
  \begin{cases}
    e^{s/\tau}\!\left(1 - e^{-T/\tau}\right), & -t_0 < s < 0, \\[3pt]
    1 - e^{-(T-s)/\tau},                      & 0 \le s < T,
  \end{cases}
  \label{eq:dark_a}
\end{equation}
and $t_0 > 0$ is the pre-gate window.
Both $T$ and $t_0$ are fixed by the readout window definition; once the
integration gate is set, $f_d^{(1)}(K)$ is fully determined by $\tau$.
Treating $s$ as uniform on $(-t_0, T)$, the charge fraction
$K = a(s)$ has the two-branch PDF obtained by Jacobian transformation.
For the branch $s\in(-t_0,0)$, $K = e^{s/\tau}(1-e^{-T/\tau})$ gives
$\mathrm{d}s/\mathrm{d}K = \tau/K$, while for $s\in[0,T)$,
$K = 1-e^{-(T-s)/\tau}$ gives $|\mathrm{d}s/\mathrm{d}K| = \tau/(1-K)$.
Combining both branches:
\begin{equation}
  f_d^{(1)}(K)
  = \frac{\tau}{t_0+T}
  \begin{cases}
    \dfrac{1}{1-K},              & 0 < K \le K_\mathrm{min}, \\[6pt]
    \dfrac{1}{K}+\dfrac{1}{1-K}, & K_\mathrm{min} < K \le K_\mathrm{max},
  \end{cases}
  \label{eq:dark_density}
\end{equation}
with $K_\mathrm{max} = 1-e^{-T/\tau}$ and
$K_\mathrm{min} = e^{-t_0/\tau}(1-e^{-T/\tau})$.
The $1/(1-K)$ branch arises from pre-gate pulses (partial tail integration);
the $1/K+1/(1-K)$ branch from in-gate pulses (partial leading-edge
integration).
\Cref{eq:dark_density} is the charge-domain formula of
\textsc{PeakOTron}~\cite{rolph_peakotron_2023}, which uses it as the
building block for charge-domain auto-convolutions.
The single-dark-pulse characteristic function is the Fourier transform of
$G^*\,f_d^{(1)}(K)$:
\begin{equation}
\begin{aligned}
  \tilde{d}_1(\omega)
  &= \int_0^{K_\mathrm{max}} e^{i\omega G^* K}\,f_d^{(1)}(K)\,\mathrm{d}K \\
  &= \frac{1}{t_0 + T}
    \int_{-t_0}^{T} e^{i\omega G^*\,a(s)}\,\mathrm{d}s.
\end{aligned}
  \label{eq:d1_cf}
\end{equation}
The second form follows by substituting $K=a(s)$ back into the first.
Evaluating \cref{eq:dark_density} in terms of the complex exponential
integral $E_1(z) = \int_z^\infty e^{-t}/t\,\mathrm{d}t$,
\begin{equation}
\begin{aligned}
  \tilde{d}_1(\omega)
  &= \frac{\tau}{t_0+T}\biggl\{
    e^{i\omega G^*}\bigl[E_1\!\bigl(i\omega G^*\,e^{-T/\tau}\bigr) \\
  &\quad - E_1(i\omega G^*)\bigr]
    + E_1\!\bigl(-i\omega G^* K_\mathrm{min}\bigr) \\
  &\quad - E_1\!\bigl(-i\omega G^* K_\mathrm{max}\bigr)
  \biggr\},
\end{aligned}
  \label{eq:d1_E1}
\end{equation}
where $E_1$ is evaluated with complex arguments using
$E_1(iy) = -\mathrm{Ci}(y) - i[\mathrm{Si}(y)-\pi/2]$ for $y > 0$.
In practice \cref{eq:d1_cf} is evaluated directly by numerical quadrature
on the frequency grid at $\mathcal{O}(N_\omega)$ cost.

\subsection{Compound-Poisson characteristic function}

Dark pulses are an independent Poisson process with rate
$r_d = \mathrm{DCR}$ over $(-t_0, T)$.
The total dark-count number is $M_d\sim\mathrm{Poisson}(\mu_d)$, with
\begin{equation}
  \mu_d = \mathrm{DCR}\,(t_0 + T).
  \label{eq:mu_dark}
\end{equation}
Since each dark pulse contributes an independent charge drawn from the mark
distribution encoded in $\tilde{d}_1$, the compound-Poisson characteristic
function of the total dark-count charge is
\begin{equation}
  \tilde{D}(\omega)
  = \exp\!\bigl[\mu_d\,\bigl(\tilde{d}_1(\omega) - 1\bigr)\bigr].
  \label{eq:D_cf}
\end{equation}
Substituting $\tilde{d}_1$ from \cref{eq:d1_cf}, this is equivalent to
the rate-integral form
$\exp\!\bigl[r_d\!\int_{-t_0}^{T}(e^{i\omega G^*\,a(s)}-1)\,\mathrm{d}s\bigr]$;
\cref{eq:D_cf} is the adopted final expression, evaluated from the
pre-computed $\tilde{d}_1(\omega_k)$ on the frequency grid.
\Cref{eq:D_cf} is the closed-form characteristic-function
counterpart of \textsc{PeakOTron}'s charge-domain auto-convolution
series~\cite{rolph_peakotron_2023}, which truncates at
$i_\mathrm{dark}^\mathrm{max}=6$ terms.
For the stated independent compound-Poisson dark-count model,
\cref{eq:D_cf} is exact at all values of $\mu_d$, including
high dark-count rates, at the same $\mathcal{O}(N_\omega)$ cost as the
light-pulse term.

\subsection{Full charge-spectrum characteristic function with dark counts}

The dark-count contribution is statistically independent of the prompt-light
contribution (independent Poisson processes); statistical independence of
additive charge contributions corresponds to pointwise multiplication of
CFs.
Adding $\tilde{D}(\omega)$~\cref{eq:D_cf} as the third factor in
\cref{eq:G_full_preview} completes the full charge-spectrum CF,
\begin{equation}
  \tilde{G}_\mathrm{full}(\omega)
  = \tilde{g}_0(\omega)
  \cdot G_N\!\bigl(\tilde{f}_\mathrm{eff}(\omega)\bigr)
  \cdot \tilde{D}(\omega),
  \label{eq:G_full_dark}
\end{equation}
where $G_N$ is the Generalised Poisson PGF~\cref{eq:gp_pgf_form},
$\tilde{f}_\mathrm{eff}$ incorporates NB afterpulsing with the
$\mathrm{Beta}(2,b)$ charge mark~\cref{eq:f_eff}, and $\tilde{D}$ is given
by \cref{eq:D_cf}.
Setting $\mathrm{DCR}=0$ (i.e.\ $\mu_d=0$) recovers $\tilde{D}=1$ and
returns the signal-only spectrum~\cref{eq:G_full}.

\Cref{eq:G_full_dark} adds one free parameter ($\mu_d$, or
equivalently $\mathrm{DCR}$) to the eight-parameter baseline model, with
$t_0$, $T$, and $\tau$ either measured from the detector pulse shape or
fixed to device-geometry defaults.
The pre-computation factorisation of Sec.~\ref{sec:reconstruction} extends
directly: the $\lambda$-independent part $\tilde{D}(\omega)$ is computed
once and stored alongside $h(\omega_k)$ and $\tilde{g}_0(\omega_k)$, so
the per-channel evaluation at any $\lambda_j$ requires only elementwise
multiplication,
\begin{equation}
  \tilde{G}_\mathrm{full}(\omega_k;\,\lambda_j)
  = \tilde{g}_0(\omega_k)\,
    e^{\lambda_j h(\omega_k)}\,
    \tilde{D}(\omega_k),
  \label{eq:G_full_precomp}
\end{equation}
at the same $\mathcal{O}(N_\omega)$ cost per channel, independent of
$k_\mathrm{max}$, $A_\mathrm{max}$, or $i_\mathrm{dark}^\mathrm{max}$.

\section{Validation}\label{sec:validation}

\subsection{Fitting procedure}

The model \cref{eq:G_full_explicit} is fitted to charge spectra
from a Hamamatsu S13360-series MPPC (PCB6)~\cite{rolph_peakotron_2023,rolph_2023_10014537},
with analysis code and fit results available at
\url{https://github.com/6yq/SiPM-charge-validation}~\cite{sipm_charge_validation},
recorded under controlled LED illumination at bias voltages 53\,V to 60\,V
in 0.5\,V steps using the measurement and data-reduction pipeline of
\textsc{PeakOTron}~\cite{rolph_peakotron_2023}.
At each voltage, $\lambda$ is varied by adjusting LED intensity to
cover the sub-1-PE to multi-PE regime, providing simultaneous
sensitivity to all model parameters.

The characteristic function \cref{eq:G_full_explicit} is evaluated
on a uniform frequency grid $\omega_k = 2\pi k/L$
($k=0,\ldots,N_\omega-1$, $L = N_\omega\,\Delta q$).
The expected count in bin $[a,b]$ is $N\!\int_a^b G(q)\,\mathrm{d}q$;
exchanging the sum $G(q)=L^{-1}\sum_k \tilde{G}_k e^{i\omega_k q}$
and the integral gives
\begin{equation*}
  \int_a^b G(q)\,\mathrm{d}q = I(b) - I(a),
\end{equation*}
where
\begin{equation}
  I(q) = \frac{\tilde{G}_0}{L}\,q
         + \frac{1}{L}\sum_{k=1}^{N_\omega-1}
           \frac{\tilde{G}_k}{i\omega_k}\,e^{i\omega_k q}.
  \label{eq:antideriv}
\end{equation}
\Cref{eq:antideriv} is exact for any bin edges.
Since $G(q)$ is real, the imaginary part of $I(b)-I(a)$ vanishes
analytically; any residual imaginary component in floating-point evaluation
is discarded as rounding noise bounded by machine epsilon times
$\|\tilde{G}\|_1$.
The fit uses the native ADC binning ($\Delta q = 1$\,ADC),
which trivially satisfies $\Delta q \ll \sigma_0$
(\cref{eq:aliasing}).
A binned extended maximum likelihood is minimised on this histogram.
The dark-count parameters $T$ and $t_0$ are set to match the
\textsc{PeakOTron} LED measurement protocol~\cite{rolph_peakotron_2023};
$\tau$ is fixed to the device pulse-decay time constant.
The fit range is $q \geq m_0 - n_\sigma\,\sigma_0$, where $n_\sigma$
is chosen to exclude readout artefacts below the pedestal and varies
with voltage: $n_\sigma = 2.5$ at 53--54\,V, $3.0$ at 54--55\,V, and
$3.5$ above 55\,V.
The upper bound is taken at the natural edge of the observed spectrum.

The likelihood and its parameter gradient are implemented in
JAX~\cite{jax2018github}; the fitter is publicly available at
\url{https://github.com/6yq/SiPM-charge-fitter}~\cite{sipm_charge_fitter}.
All LED illumination levels from the PeakOTron data release~\cite{rolph_2023_10014537}
are used at each voltage.
Parameters are expressed in unconstrained form (log or logit transforms)
to keep the Gaussian approximation valid.
Optimisation uses L-BFGS with a zoom line search and 16 random
initialisations, 
from which the run with the best final likelihood is selected as the
global optimum.
The inverse Hessian from JAX automatic differentiation provides parameter
uncertainties and correlations without finite differences.
The fitter runs on CPU, yet JAX compiles to GPU without code changes, 
enabling acceleration for large-scale surveys.

\subsection{Model selection by BIC}
\label{sec:bic}

Afterpulsing and dark counts each introduce additional free parameters
whose identifiability depends on the gain-to-noise ratio $G^*/\sigma_0$.
We compare four nested models:
\textbf{M1}~(baseline, $k=7$): cross-talk only, no afterpulsing, no dark counts;
\textbf{M2}~(AP only, $k=9$): adds $\rho$ and $b$;
\textbf{M3}~(DCR only, $k=8$): adds $\mu_d$;
\textbf{M4}~(full, $k=10$): all effects.
All models use the GP cross-talk CF~\cref{eq:crosstalk_cf};
M1 and M3 set $\rho=0$ in \cref{eq:G_full_explicit},
M2 and M4 enable the full NB afterpulsing term;
M3 and M4 add the dark-count factor $\tilde{D}$~\cref{eq:D_cf}.
The Bayesian Information Criterion~\cite{schwarz_estimating_1978},
\begin{equation}
  \mathrm{BIC} = -2\ln\hat{\mathcal{L}} + k\ln N,
  \label{eq:bic}
\end{equation}
where $\hat{\mathcal{L}}$ is the maximised likelihood, $k$ the number
of free parameters, and $N$ the total number of fitted events, penalises each
additional parameter by $\tfrac{1}{2}\ln N \approx 6.4$--$6.7$.
Table~\ref{tab:bic} shows BIC values for all four models at five
representative voltages.

\begin{table}[!htbp]
\centering
\caption{BIC comparison of four nested models.
  M1: cross-talk only, no AP, no DCR ($k=7$); M2: AP only ($k=9$);
  M3: DCR only ($k=8$); M4: AP and DCR ($k=10$).
  $N_\text{ev}$ is the total number of fitted events used in the BIC
  penalty~\cref{eq:bic}.
  Bold marks the BIC-selected model at each voltage.}
\label{tab:bic}
\small
\begin{tabular}{lcrrrr}
\hline
$V_\text{bias}$ & $N_\text{ev}$ ($10^3$) &
  \multicolumn{1}{c}{M1} & \multicolumn{1}{c}{M2} &
  \multicolumn{1}{c}{M3} & \multicolumn{1}{c}{M4} \\
\hline
53.0\,V & 334  & \textbf{2207.8} & 2225.6 & 2217.1 & 2238.4 \\
53.5\,V & 577  & \textbf{3797.5} & 3824.0 & 3804.8 & 3831.0 \\
54.0\,V & 390  & 4336.2 & 4264.5 & \textbf{3566.7} & 3591.4 \\
54.5\,V & 332  & 7457.0 & 6219.3 & 3831.7 & \textbf{3796.8} \\
55.0\,V & 349  & 15719.6 & 10726.6 & 4693.3 & \textbf{4471.1} \\
\hline
\end{tabular}
\end{table}

The BIC selects different models in three voltage regimes.
At 53--53.5\,V ($G^*/\sigma_0\approx4$--$5.5$) neither the
$K\in(0,1)$ valley (dark counts) nor the $K\in(n,n+1)$ valleys for
$n\geq1$ (afterpulses) carry independent information: both M2 and M3
are disfavoured ($\Delta\mathrm{BIC}\geq+9$), and M1 is selected.
At 54\,V ($G^*/\sigma_0\approx7.2$) the 0-PE valley just resolves
and DCR becomes identifiable ($\Delta\mathrm{BIC}(\mathrm{M1}\to\mathrm{M3})=+769.5$),
but the inter-PE valleys still cannot constrain $\rho$ and $b$
($\Delta\mathrm{BIC}(\mathrm{M3}\to\mathrm{M4})=+24.7$); M3 is selected.
At $\geq54.5$\,V ($G^*/\sigma_0\geq8.8$) the inter-PE valleys widen
and afterpulsing is identified ($\Delta\mathrm{BIC}(\mathrm{M3}\to\mathrm{M4})=-34.9$
at 54.5\,V, reaching $-222$ at 55\,V); M4 is selected for all higher voltages.

\subsection{Fit results at representative voltages}

\begin{figure}[!htbp]
  \includegraphics[width=0.9\hsize]{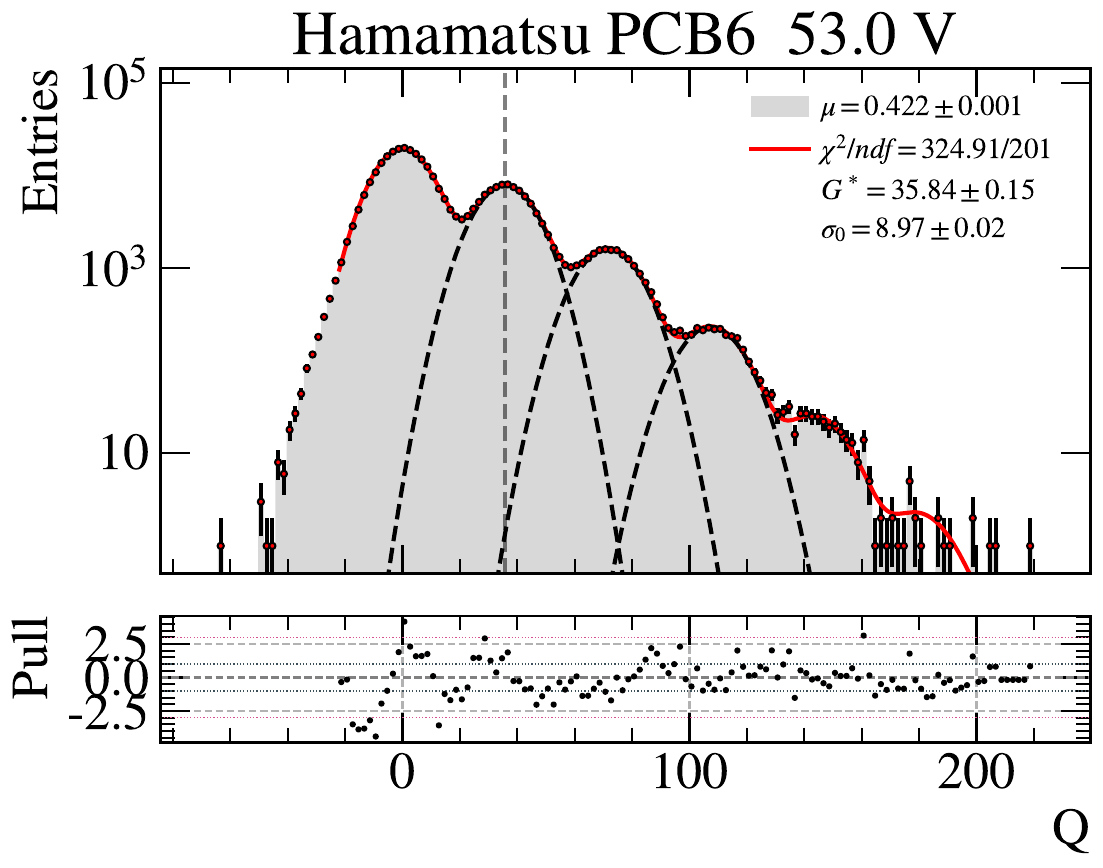}
  \caption{Charge spectrum and BIC-selected fit (M1: no AP, no DCR)
    for PCB6 at 53\,V ($\lambda=0.422$).
    Dashed curves show individual $n$-PE components.
    Pull panel (lower): normalised residuals ($\pm 2.5$ bands);
    deviations in the 0-PE region reflect non-Gaussian pedestal tails
    present in all four models.}
  \label{fig:fit53}
\end{figure}

\begin{figure}[!htbp]
  \includegraphics[width=0.9\hsize]{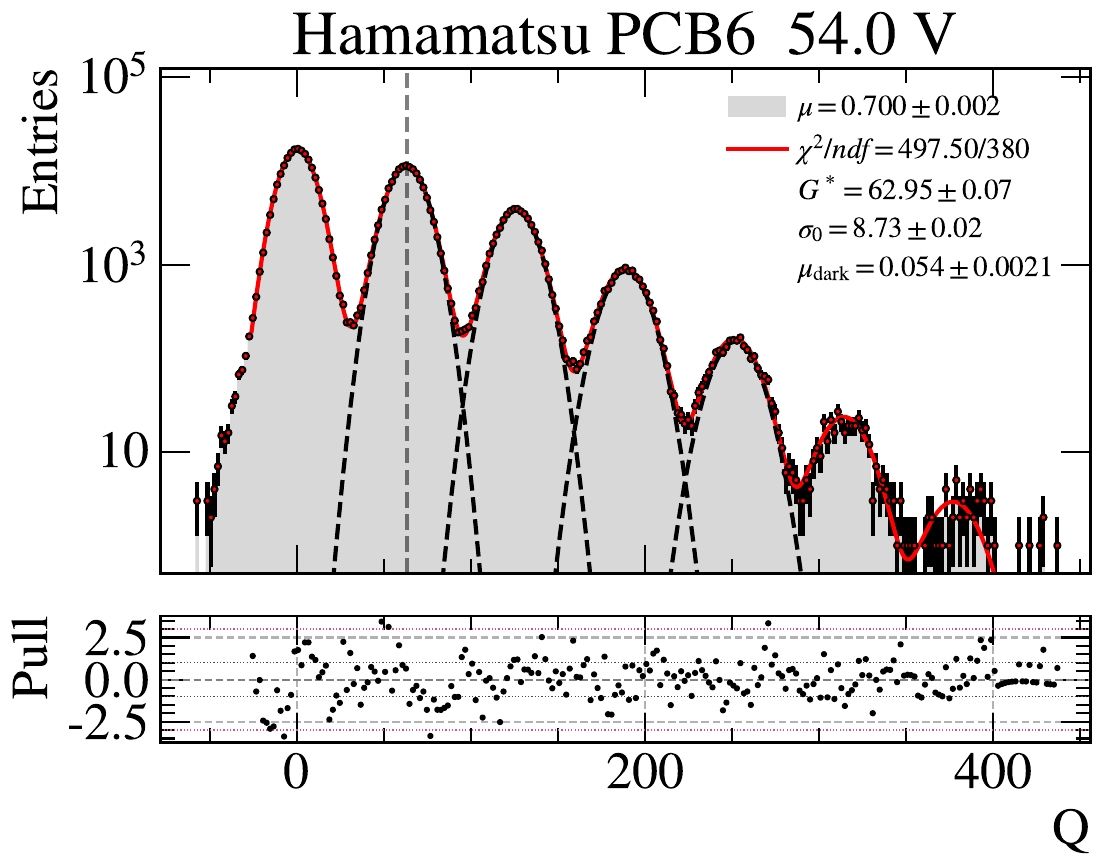}
  \caption{BIC-selected fit (M3: DCR only, no AP) for PCB6 at 54\,V
    ($\lambda=0.700$).
    The $K\in(0,1)$ valley is clearly resolved; the $K\in(1,2)$ valley
    carries no afterpulse structure, consistent with
    $\Delta\mathrm{BIC}(\mathrm{M3}\to\mathrm{M4})=+24.7$.}
  \label{fig:fit54}
\end{figure}

\begin{figure}[!htbp]
  \includegraphics[width=0.9\hsize]{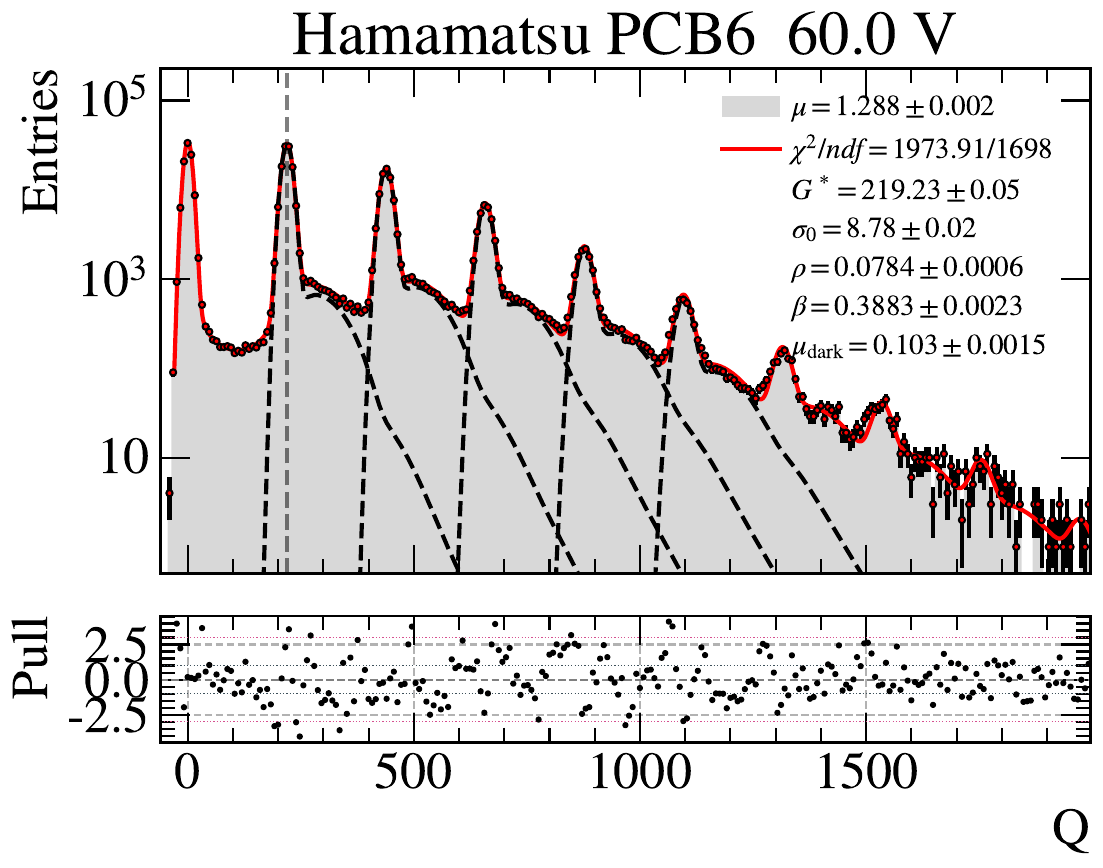}
  \caption{BIC-selected fit (M4: full model) for PCB6 at 60\,V
    ($\lambda=1.288$).
    Six PE peaks are resolved; afterpulse charge fills all inter-PE
    valleys and the dark-count continuum fills $K\in(0,1)$.}
  \label{fig:fit60}
\end{figure}

\begin{figure}[!htbp]
  \includegraphics[width=0.9\hsize]{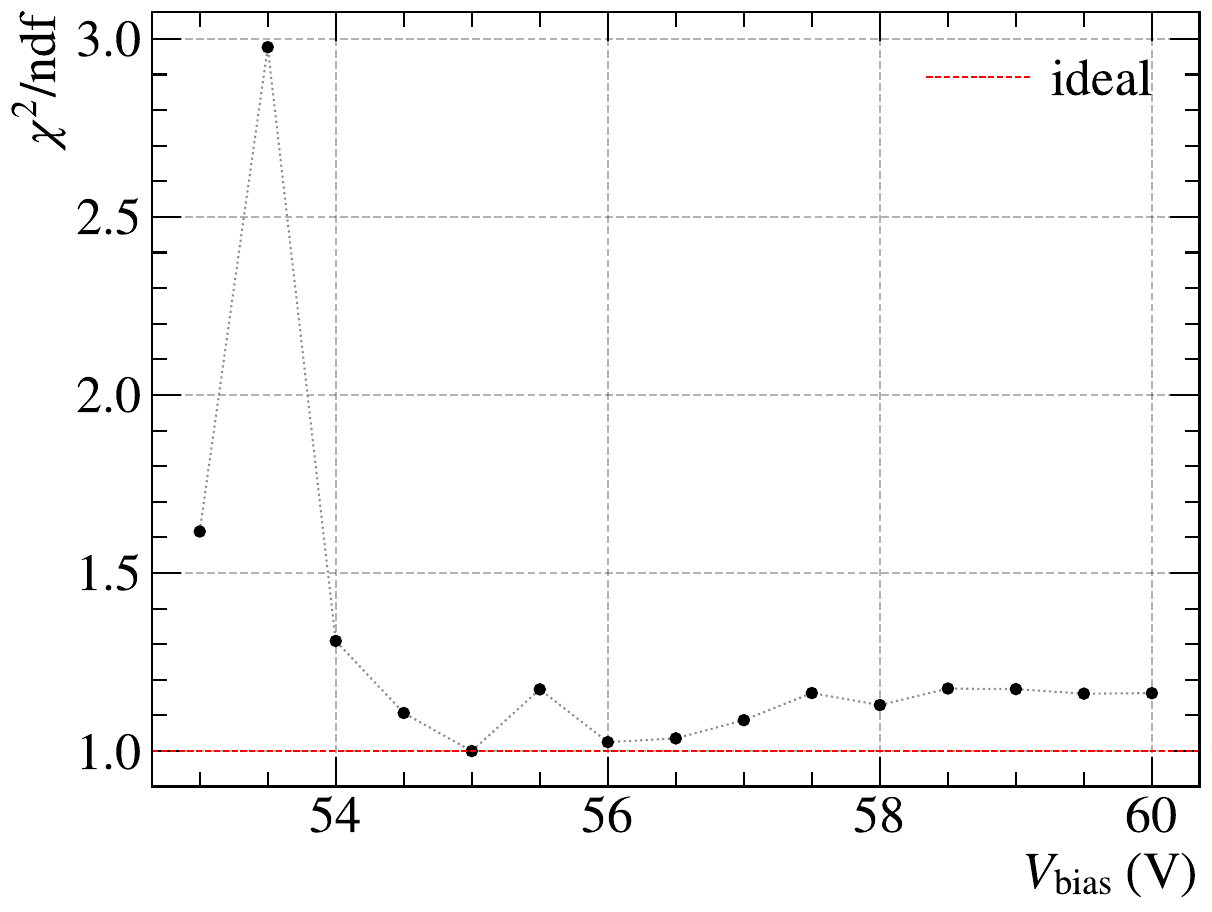}
  \caption{$\chi^2/\text{ndf}$ versus bias voltage for the
    BIC-selected model.
    The dashed line marks the ideal value of 1.
    Elevated values at 53--53.5\,V arise from non-Gaussian pedestal
    tails and are independent of model choice.}
  \label{fig:chisq}
\end{figure}

Figures~\ref{fig:fit53}--\ref{fig:fit60} show BIC-selected fits at
three voltages spanning the selection transitions:
53\,V (M1), 54\,V (M3), and 60\,V (M4).
Figure~\ref{fig:chisq} shows $\chi^2/\text{ndf}$ for the BIC-selected
model across all voltages.
The elevated values at 53--53.5\,V are driven by the Gaussian pedestal
not capturing non-Gaussian electronic noise tails in the 0-PE region;
this systematic is independent of model choice and resolves as
$G^*/\sigma_0$ grows above 54\,V.

\begin{table}[!htbp]
\centering
\caption{Representative fitted physical parameters for the BIC-selected models
  at three voltages.
  $G^*$ is the effective gain in ADC units ($G^*=\alpha\beta$).
  Dashes indicate parameters not included in the selected model.}
\label{tab:params}
\small
\begin{tabular}{lcccc}
\hline
$V_\text{bias}$ & $G^*$ (ADC) & $\theta$ & $\rho$ & $\mu_d$ \\
\hline
53.0\,V (M1) & $35.8\pm0.2$   & $0.0043\pm0.0013$ & --- & --- \\
54.0\,V (M3) & $62.9\pm0.1$   & $0.0061\pm0.0012$ & --- & $0.054\pm0.002$ \\
60.0\,V (M4) & $219.2\pm0.1$  & $0.0470\pm0.0012$ & $0.0784\pm0.0006$ & $0.104\pm0.002$ \\
\hline
\end{tabular}
\end{table}

\subsection{Parameter correlations and identifiability}
\label{sec:corr}

\begin{figure}[!htbp]
  \includegraphics[width=0.9\hsize]{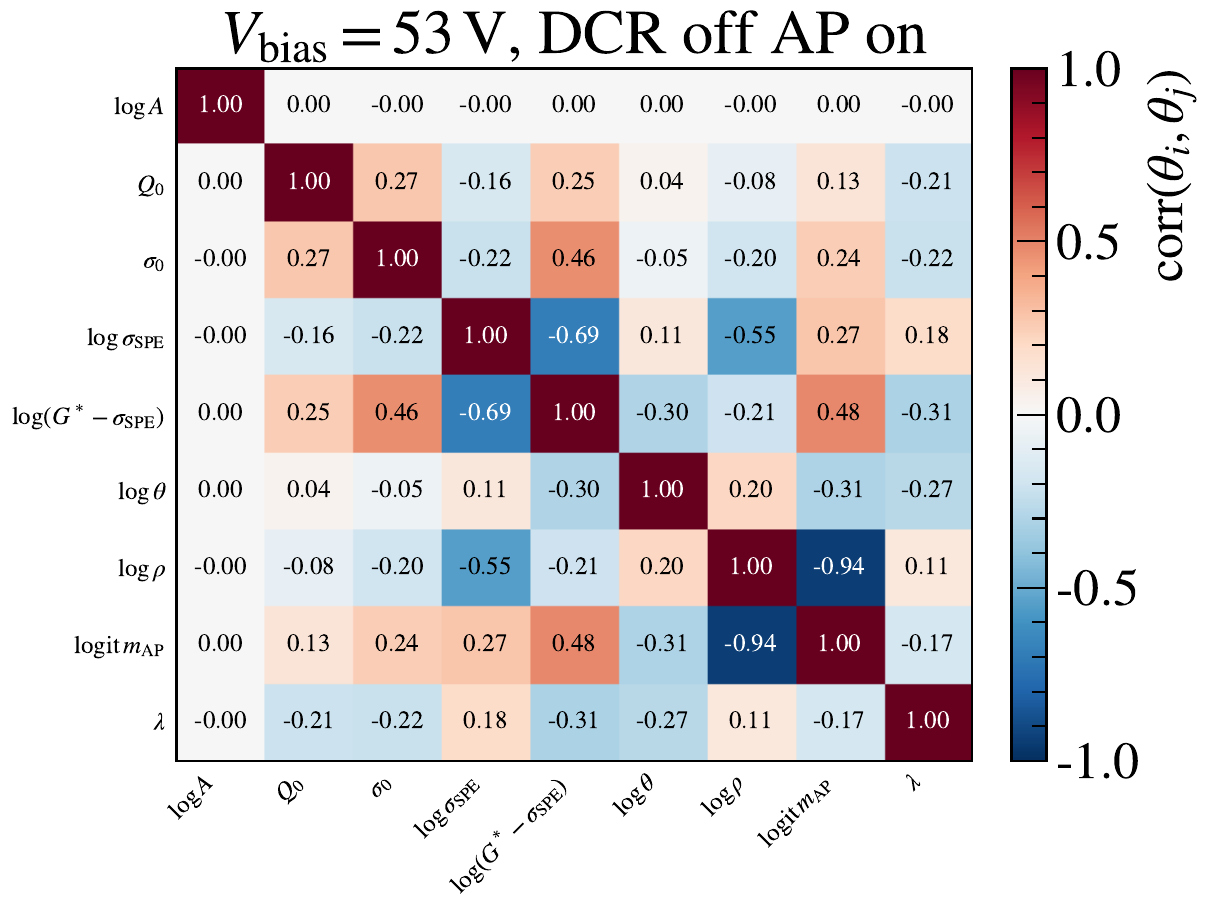}
  \caption{MLE parameter correlation matrix at 53\,V (forced M2 fit,
    9 parameters).
    The dominant feature is the $\log\rho$--$\text{logit}\,m_\text{AP}$
    anti-correlation ($r=-0.94$): with $G^*/\sigma_0\approx4$,
    inter-PE valleys cannot independently constrain the afterpulse rate
    $\rho$ and charge fraction $m_\text{AP}$, driving BIC to reject AP.
    The $\log\sigma_\text{SPE}$--$\log(G^*-\sigma_\text{SPE})$
    anti-correlation ($r=-0.69$) is intrinsic to the two-parameter
    Gamma SPE.}
  \label{fig:corr53}
\end{figure}

\begin{figure}[!htbp]
  \includegraphics[width=0.9\hsize]{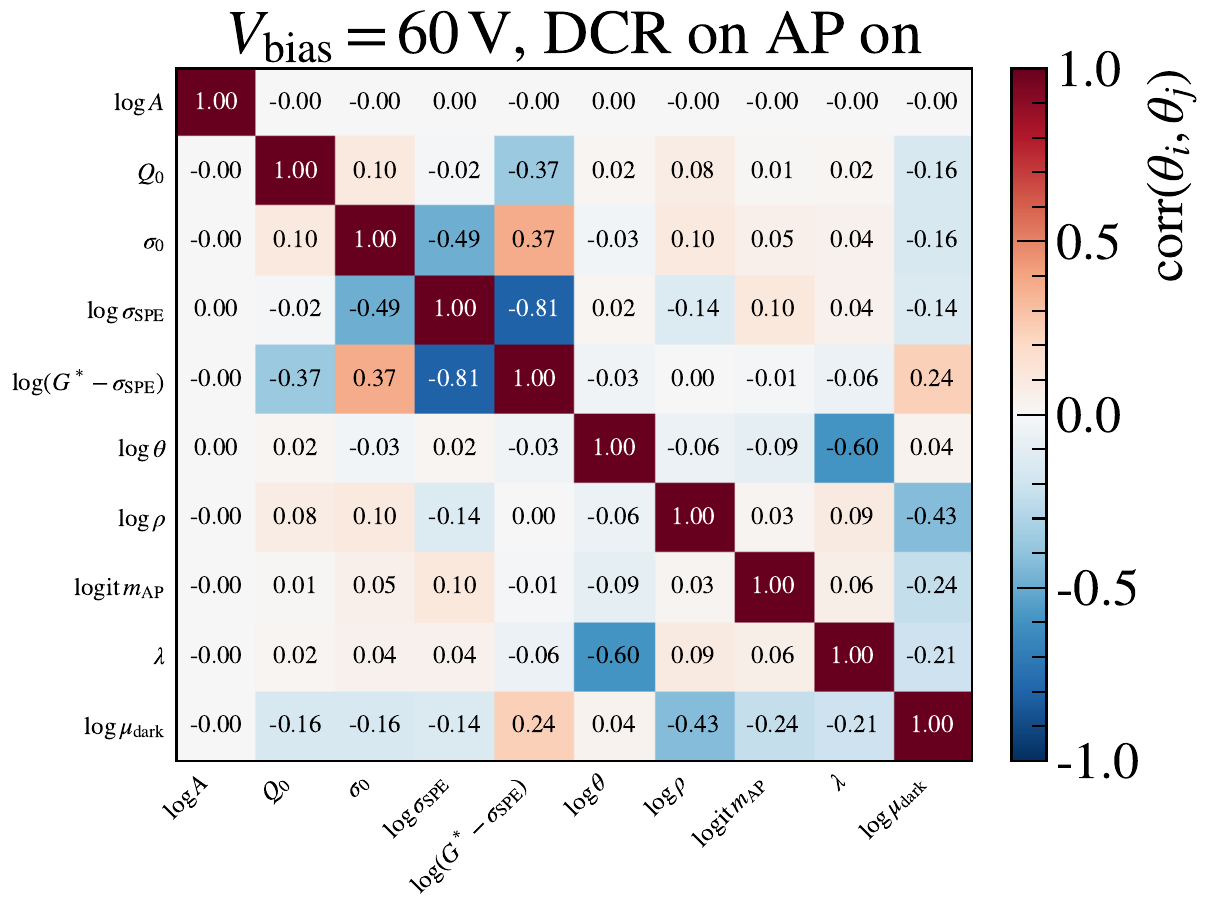}
  \caption{MLE parameter correlation matrix at 60\,V (M4, 10
    parameters).
    The $\rho$--$m_\text{AP}$ degeneracy has resolved ($r\approx0.03$).
    Notable features: $\log\theta$--$\lambda$ anti-correlation
    ($r=-0.60$) from cross-talk inflating apparent PE count, and
    $\log\rho$--$\log\mu_d$ anti-correlation ($r=-0.43$) from partial
    overlap of afterpulse and dark-count charge in $K\in(0,1)$.}
  \label{fig:corr60}
\end{figure}

Figures~\ref{fig:corr53} and~\ref{fig:corr60} show MLE parameter
correlation matrices at 53\,V and 60\,V.
At 53\,V the forced M2 (AP-on, DCR-off) fit is shown to illustrate
the $\rho$--$b$ degeneracy driving BIC to prefer M1.
The parameterisation uses log or logit transforms:
$\log A$ ($A$ = Poisson mean of the total observed event count in the
extended likelihood), $Q_0\equiv m_0$ (pedestal mean in ADC units),
$\sigma_0$, $\log\sigma_\text{SPE}$,
$\log(G^*-\sigma_\text{SPE})$, $\log\theta$, $\log\rho$,
$\text{logit}\,m_\text{AP}$ ($\equiv\text{logit}\,\frac{2}{2+b}$),
$\lambda$, and $\log\mu_d$ (when DCR is included).

Four correlations are notable.

\emph{Afterpulse degeneracy at low voltage.}
As Fig.~\ref{fig:corr53} shows, $(\log\rho,\,\text{logit}\,m_\text{AP})$
is nearly degenerate ($r=-0.94$) at 53\,V: the inter-PE valleys at
$G^*/\sigma_0\approx4$ cannot independently constrain $\rho$ and
$m_\text{AP}$, so BIC rejects the two-parameter extension.
At 60\,V ($r\approx0.03$) both are well constrained.

\emph{Intrinsic SPE shape correlation.}
$(\log\sigma_\text{SPE},\,\log(G^*-\sigma_\text{SPE}))$ shows a
persistent anti-correlation ($r=-0.69$ at 53\,V, $-0.81$ at 60\,V),
intrinsic to the two-parameter Gamma representation; not an
identifiability problem.

\emph{Cross-talk--$\lambda$ correlation.}
$(\log\theta,\,\lambda)$ develops $r=-0.60$ at 60\,V: higher cross-talk
adds secondary discharges that inflate the apparent charge, compensated
by lower $\lambda$.

\emph{Afterpulse--dark-count correlation.}
$(\log\rho,\,\log\mu_d)$ shows $r=-0.43$ at 60\,V from partial overlap
of the two effects in $K\in(0,1)$.
They remain resolvable because dark counts are confined to $K\in(0,1)$
while afterpulses appear in all $K\in(n,n+1)$ valleys.

\subsection{Parameter evolution with bias voltage}

\begin{figure}[!htbp]
  \includegraphics[width=0.9\hsize]{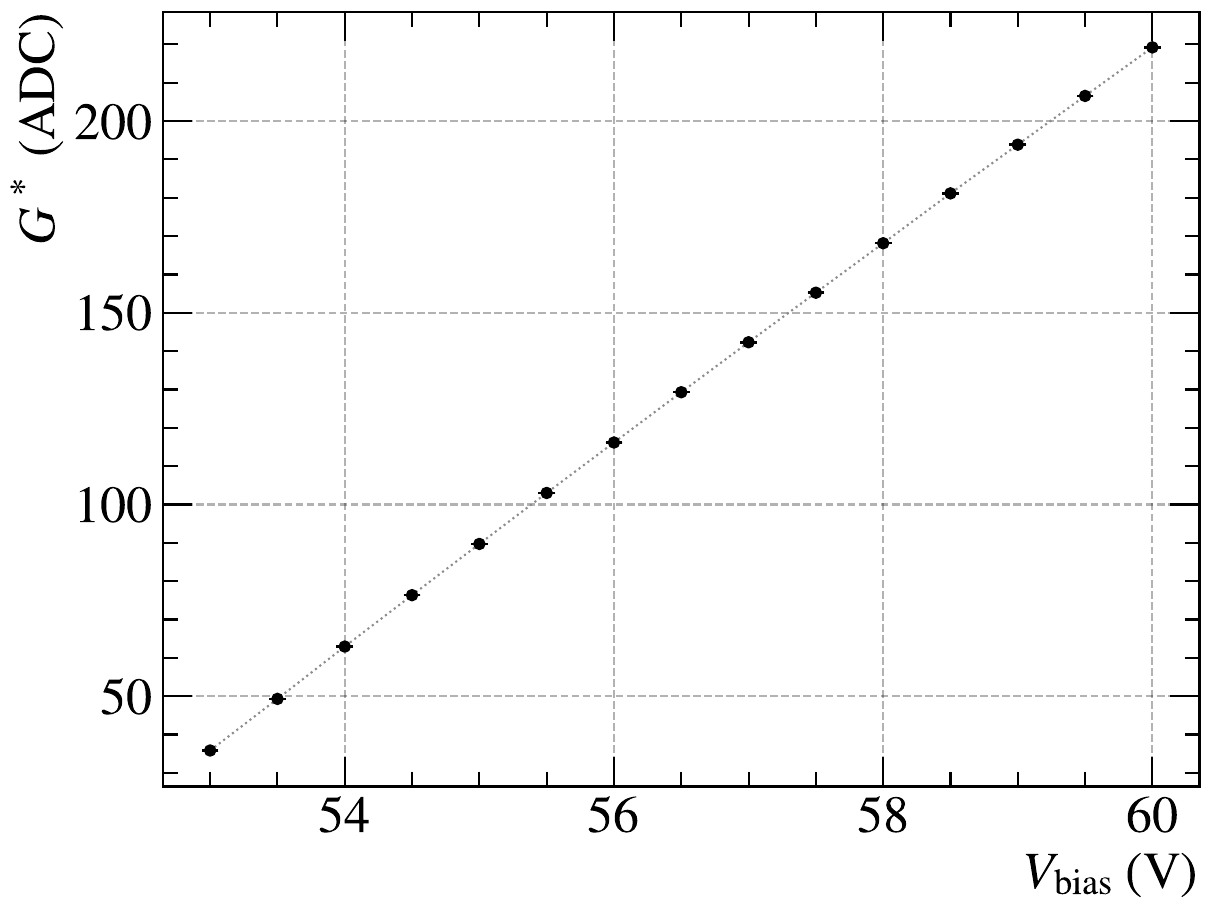}
  \caption{Effective gain $G^*$ versus bias voltage from
    BIC-selected fits.
    The linear dependence is consistent with
    $G^*\propto(V_\text{bias}-V_\text{bd})$.}
  \label{fig:gain}
\end{figure}

\begin{figure}[!htbp]
  \includegraphics[width=0.9\hsize]{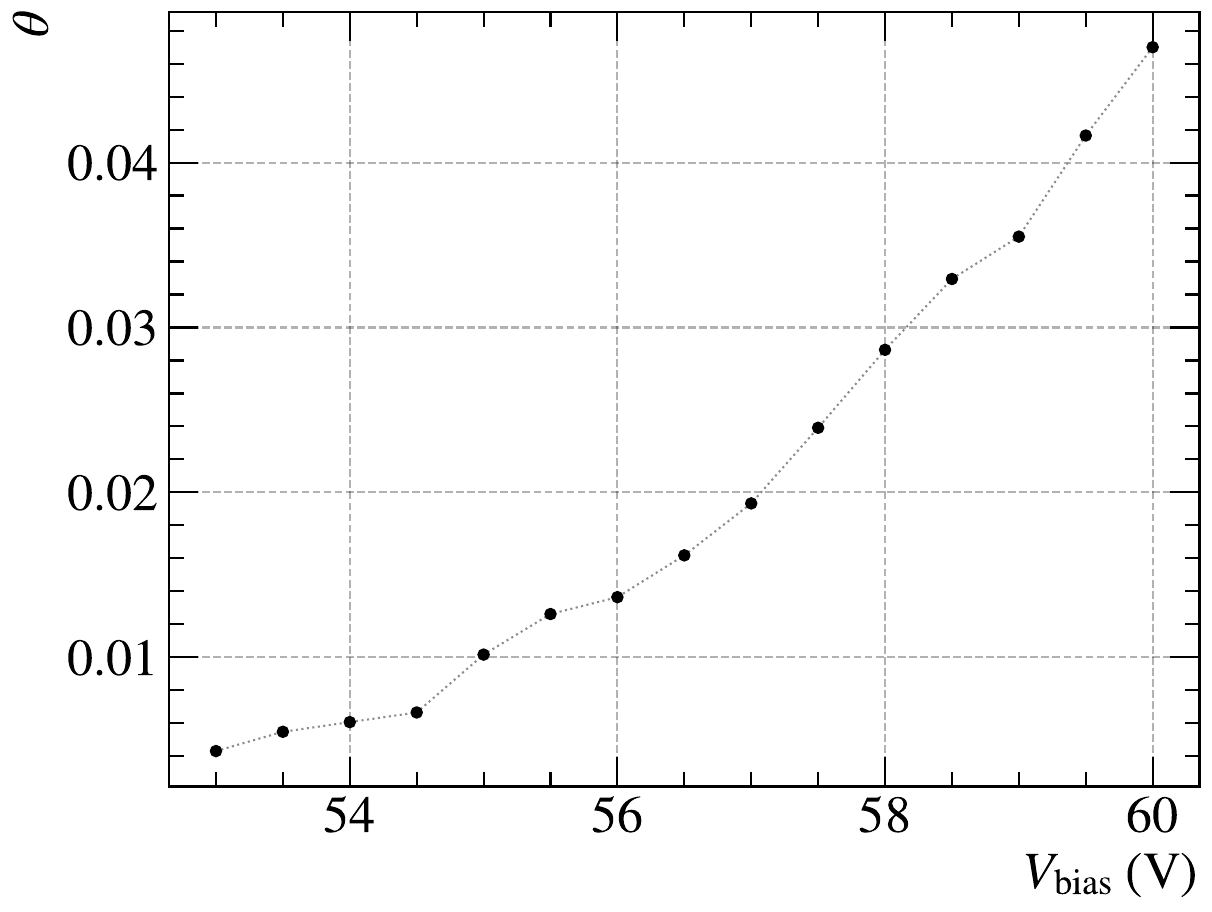}
  \caption{Cross-talk parameter $\theta$ versus bias voltage.
    Super-linear growth reflects increasing hot-carrier luminescence
    yield with overvoltage.}
  \label{fig:theta}
\end{figure}

\begin{figure}[!htbp]
  \includegraphics[width=0.9\hsize]{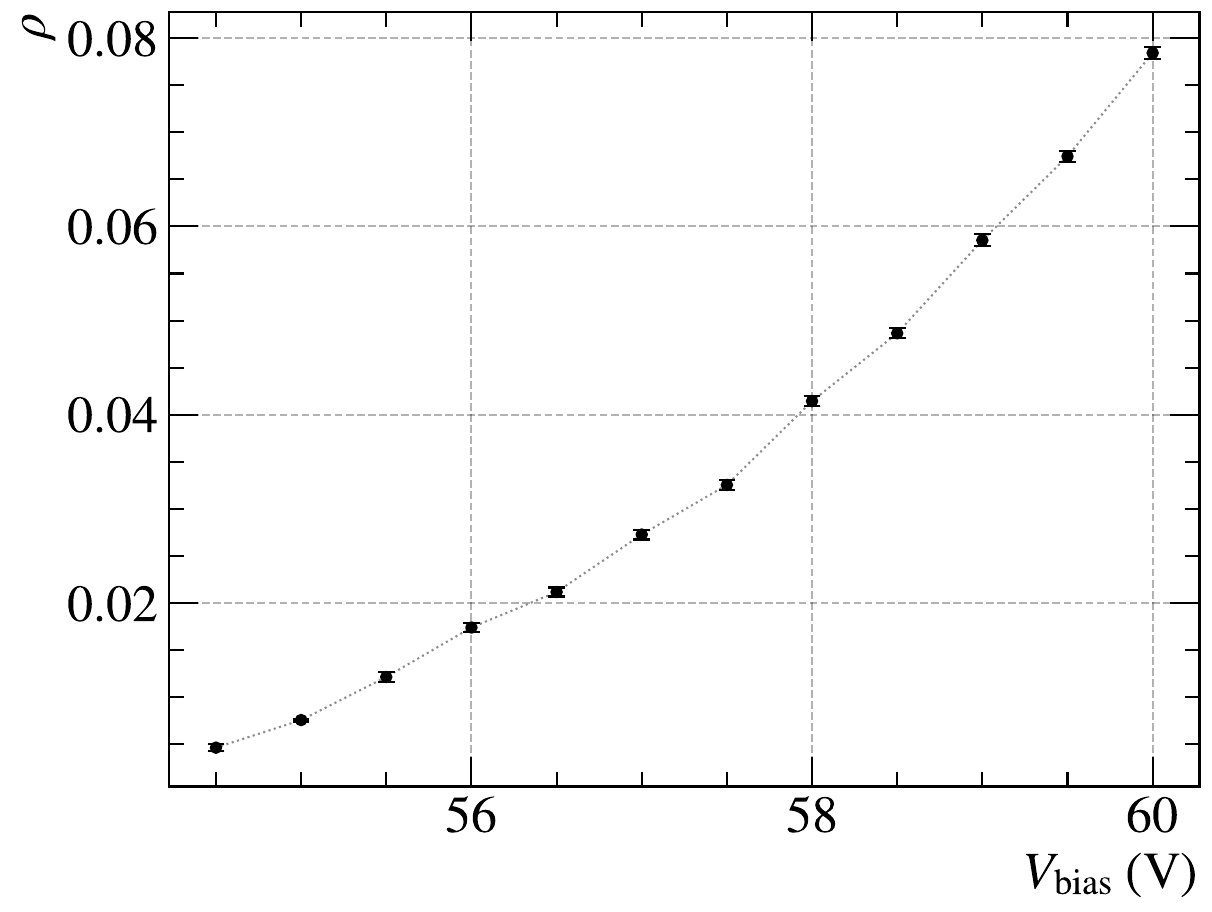}
  \caption{Afterpulse rate $\rho$ versus bias voltage.
    Data points start at 54.5\,V where BIC first selects a model
    including afterpulsing (M4).}
  \label{fig:rho}
\end{figure}

\begin{figure}[!htbp]
  \includegraphics[width=0.9\hsize]{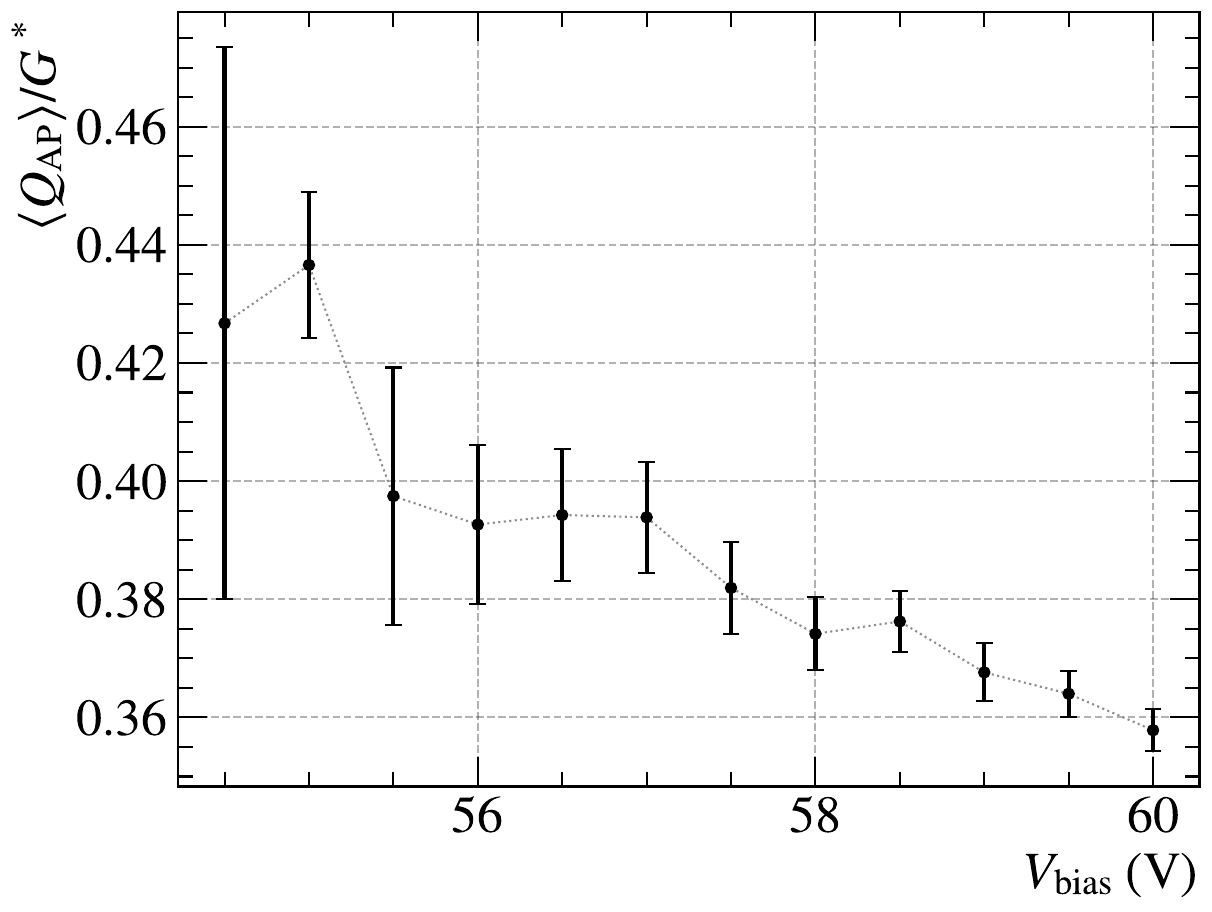}
  \caption{Mean afterpulse charge fraction
    $\langle Q_\text{AP}\rangle/G^* = 2/(2+b)$ versus bias voltage
    (from 54.5\,V onwards).
    The fraction stabilises near 0.43, corresponding to $b\approx2.7$
    and indicating typical afterpulse arrivals at $\approx43\%$ SPAD
    recovery.}
  \label{fig:map}
\end{figure}

Figures~\ref{fig:gain}--\ref{fig:map} show four physics parameters
from the BIC-selected fits as functions of $V_\text{bias}$.
The gain $G^*$ (Fig.~\ref{fig:gain}) is linear in $V_\text{bias}$,
consistent with $G^*\propto(V_\text{bias}-V_\text{bd})$.
The cross-talk parameter $\theta$ (Fig.~\ref{fig:theta}) grows
super-linearly above breakdown, reflecting increasing hot-carrier
luminescence yield with overvoltage.
The afterpulse rate $\rho$ (Fig.~\ref{fig:rho}) is not identified
below 54.5\,V (BIC selects AP-off models; Sec.~\ref{sec:bic}) and
rises monotonically from 54.5\,V onwards.
Similarly, $\langle Q_\text{AP}\rangle/G^*$ (Fig.~\ref{fig:map}) is
first constrained from 54.5\,V and stabilises near 0.43, indicating
typical afterpulse arrivals at $\approx43\%$ SPAD recovery.
The DCR parameter $\mu_d$ is first identified at 54\,V
(Sec.~\ref{sec:bic}) and rises monotonically to $0.103\pm0.002$ at
60\,V.
Stability of $\rho$ and $b$ across LED intensities at fixed bias voltage
was not tested; each voltage uses all available LED data but provides
a single joint fit rather than per-intensity cross-checks.
The results reported here are specific to the Hamamatsu S13360-series
PCB6 device at room temperature; parameter values will differ for
other SiPM types, operating temperatures, or overvoltages outside
the 53--60\,V range surveyed.

\section{Application to Event Reconstruction}\label{sec:reconstruction}

The analytical form of $\tilde{G}$ in \cref{eq:G_full_explicit} admits
closed-form evaluation for event-level energy reconstruction.
We extend the joint charge-time probability density
of~\cite{tao_reconstruction_2025} to the closed-form charge model \cref{eq:G_full_explicit} and
show that the $(1+\partial_\lambda)$ operator has an explicit closed form
within this framework.

\subsection{Joint charge-time probability density}

Let $R_j(t;\vec{r})$ denote the expected PE rate at the $j$-th SiPM at
time $t$ for a vertex at position $\vec{r}$, within a readout window
$[\underline{T}, \bar{T}]$.
Consider two events in the sample space:
$\mathcal{A}_T$: no PE arrives before $T$, and total charge from
$\int_T^{\bar{T}} R_j\,\mathrm{d}t$ is collected; and
$\mathcal{A}_{T+\Delta T}$: no PE arrives before $T + \Delta T$, with
total charge from $\int_{T+\Delta T}^{\bar{T}} R_j\,\mathrm{d}t$.
Since $\mathcal{A}_{T+\Delta T} \subset \mathcal{A}_T$, the difference
$\mathcal{A}_T \setminus \mathcal{A}_{T+\Delta T}$ is the event that the
first PE arrives in $[T, T+\Delta T)$.
Writing $\lambda_<(T) = \int_{\underline{T}}^T R_j\,\mathrm{d}t$ and
$\lambda_>(T) = \int_T^{\bar{T}} R_j\,\mathrm{d}t$,

\begin{equation*}
\begin{aligned}
  P(\mathcal{A}_T) &= e^{-\lambda_<(T)}\,G\!\bigl(Q;\lambda_>(T)\bigr), \\
  P(T,Q)\,\Delta T
  &= P(\mathcal{A}_T) - P(\mathcal{A}_{T+\Delta T}) \\
  &= -\frac{\mathrm{d}}{\mathrm{d}T}\,P(\mathcal{A}_T)\,\Delta T,
\end{aligned}
\end{equation*}
where $G(Q;\lambda)$ denotes the charge PDF \cref{eq:Gsig} evaluated at mean
PE count $\lambda$.
Differentiating with the chain rule ($\mathrm{d}\lambda_</\mathrm{d}T =
R_j(T)$, $\mathrm{d}\lambda_>/\mathrm{d}T = -R_j(T)$),
\begin{equation}
  P(T, Q) = e^{-\lambda_<(T)}\,R_j(T)\,
  \bigl(1 + \partial_\lambda\bigr)\,G\!\bigl(Q;\lambda\bigr)
  \Big|_{\lambda = \lambda_>(T)},
  \label{eq:TQ}
\end{equation}
where $(1+\partial_\lambda)G$ groups the two terms from differentiating
the exponential prefactor and the charge distribution, and $G$ is the
full charge PDF \cref{eq:G_full_explicit}.
The formula \cref{eq:TQ} follows that of~\cite{tao_reconstruction_2025},
derived for a Tweedie charge response; $G$ here replaces the Tweedie
model to cover cross-talk and afterpulsing.

\subsection{Analytic evaluation of $(1+\partial_\lambda)G$}

The operator $(1+\partial_\lambda)$ acts simply on $\tilde{G}$:
\begin{equation}
  \bigl(1+\partial_\lambda\bigr)\tilde{G}(\omega;\lambda)
  = \tilde{G}(\omega;\lambda)\cdot T\!\bigl(\tilde{f}_{\mathrm{eff}}(\omega)\bigr),
  \label{eq:1plus_dlambda}
\end{equation}
because $\partial_\lambda \tilde{G} = \tilde{G}\cdot(T(\tilde{f}_\text{eff})-1)$
from the form \cref{eq:G_full}.
For $\theta = 0$ (Poisson), $T(s) = s$, so
$(1+\partial_\lambda)\tilde{G} = \tilde{G}\cdot\tilde{f}_\text{eff}$,
which in position space is the convolution $G * f_\text{eff}$: the
likelihood weight equals the charge spectrum convolved with the
single-cluster response.
For $\theta > 0$ the explicit expression for $T$ from
\cref{eq:T_lambertW} gives
\begin{equation}
  T\!\bigl(\tilde{f}_{\mathrm{eff}}(\omega)\bigr)
  = -\frac{1}{\theta}\,
  W_0\!\!\left(-\theta\,e^{-\theta}\,\tilde{f}_{\mathrm{eff}}(\omega)\right),
  \label{eq:T_feff}
\end{equation}
so $(1+\partial_\lambda)G(Q;\lambda)$ is still computable by inverse FFT of
$\tilde{G}\cdot T(\tilde{f}_\text{eff})$, requiring no additional
numerical quadrature beyond the FFT already used for $G$ itself.

\subsection{Full detector log-likelihood}

Denoting the set of hit channels (at least one PE detected) as
$\mathcal{H}$ and writing $\bar\lambda_j \equiv
\int_{\underline{T}}^{\bar{T}}\!R_j(t;\vec{r})\,\mathrm{d}t$ and
$\lambda_{j>}(T)\equiv\int_T^{\bar{T}}\!R_j(t;\vec{r})\,\mathrm{d}t$,
the event log-likelihood at vertex position $\vec{r}$ is
\begin{equation}
\begin{aligned}
  \log L(\vec{r})
  &= \sum_{j \in \mathcal{H}}\!
    \Bigl[
      -\bar\lambda_j
      + \log R_j(T_j;\vec{r}) \\
  &\quad + \log\!\bigl(
        (1+\partial_\lambda)\,G(Q_j;\lambda)
        \bigr|_{\lambda = \lambda_{j>}(T_j)}
      \bigr)
    \Bigr] \\
  &\quad - \sum_{j \notin \mathcal{H}} \bar\lambda_j.
\end{aligned}
\label{eq:logL}
\end{equation}
All three terms are analytically accessible: the rate integrals
$\bar\lambda_j$ follow from the light-yield model, $\log R_j(T_j)$
is evaluated at the observed first-hit time, and
$(1+\partial_\lambda)G$ is evaluated via \cref{eq:1plus_dlambda}--\cref{eq:T_feff}.
\Cref{eq:logL} generalises the time-only likelihood used in
previous time-only reconstruction analyses by incorporating the measured
charge $Q_j$ at each channel
through the full analytical charge response model
\cref{eq:G_full_explicit}, without introducing any additional numerical
integration step.

\subsection{Likelihood scan acceleration}
\label{sec:recon_demo}

The characteristic function \cref{eq:G_full_explicit} factorises the
$\lambda$-dependence cleanly.
Writing
\begin{equation}
  h(\omega) \equiv
  -\frac{1}{\theta}\,W_0\!\!\left(-\theta\,e^{-\theta}\,
    \tilde{f}_{\mathrm{eff}}(\omega)\right) - 1,
  \label{eq:h_omega}
\end{equation}
which depends only on the calibrated single-avalanche response parameters
$(\alpha, \beta, \theta, \rho, b)$ and not on $\lambda$, the baseline
CF becomes
\begin{equation}
  \tilde{G}(\omega;\lambda)
  = \tilde{g}_0(\omega)\,e^{\lambda\,h(\omega)}.
  \label{eq:G_factored}
\end{equation}
When the dark-count term $\tilde{D}(\omega)$~\cref{eq:D_cf} is
included, it is pre-computed once from the calibrated parameters
$(r_d, \tau, t_0, T)$ and enters as an additional multiplicative
factor independent of $\lambda$:
$\tilde{G}_\mathrm{full}(\omega;\lambda) =
\tilde{g}_0(\omega)\,e^{\lambda h(\omega)}\,\tilde{D}(\omega)$.
After calibration, $h(\omega_k)$ and $\tilde{D}(\omega_k)$ are
pre-computed once on the $N_\omega$-point frequency grid.
For each channel $j$ with expected PE count $\lambda_j$, the
per-channel likelihood is evaluated as:
\begin{enumerate}
  \item[(i)] Form $\tilde{G}_\mathrm{full}(\omega_k;\lambda_j) =
    \tilde{g}_0(\omega_k)\,e^{\lambda_j h(\omega_k)}\,\tilde{D}(\omega_k)$
    elementwise, $O(N_\omega)$.
  \item[(ii)] Evaluate $\log\!\bigl[(1+\partial_\lambda)G(Q_j;\lambda_j)\bigr]$
    via the Fourier series
    \begin{equation}
      G(Q_j;\lambda_j)
      = \frac{1}{N_\omega\,\Delta q}
        \sum_{k=0}^{N_\omega-1}
        \tilde{G}(\omega_k;\lambda_j)\,e^{i\omega_k Q_j}
      \label{eq:fs_eval}
    \end{equation}
    with $\tilde{G}(\omega_k;\lambda_j)\to
    \tilde{G}(\omega_k;\lambda_j)\cdot T(\tilde{f}_\mathrm{eff}(\omega_k))$
    \cref{eq:1plus_dlambda} at $\lambda_j = \lambda_{j>}(T_j)$.
    Because each channel $j$ has a distinct observed charge $Q_j$,
    these evaluations at non-uniform points $\{Q_j\}$ on a uniform
    frequency grid are naturally handled by a type-2
    NUFFT~\cite{barnett_parallel_2019}, completing the sum for all
    $M$ hit channels simultaneously in $O(N_\omega\log N_\omega + M)$,
    with aliasing error bounded by \cref{eq:aliasing}.
  \item[(iii)] For unfired channels ($j\notin\mathcal{H}$), add
    $-\bar\lambda_j$ to the log-likelihood, with no charge-model
    evaluation required.
\end{enumerate}

This contrasts with histogram-based approaches, which store
$k_\text{max}$ pre-convolved histograms $h_n(q) = (g_0 * f^{*n})(q)$
for $n = 0, \ldots, k_\text{max}$, each of length $N_q$; with
cross-talk and afterpulsing the table is two-dimensional (indexed by
$n$ and total afterpulse count $A$), of size
$O(k_\text{max} \times A_\text{max} \times N_q)$, and must be
recomputed when $\lambda$ changes.
At $\lambda=10$\,PE one needs $k_\text{max}\sim 30$--$50$, giving
$O(50\times A_\text{max}\times N_q)$ entries per recomputation.
The present approach stores only the $O(N_\omega)$ array $h(\omega_k)$
\cref{eq:h_omega}, independent of $k_\text{max}$, $A_\text{max}$, or
the $\lambda$ range.

Per-channel likelihood evaluation at any $\lambda_j$ requires a single
elementwise multiply and one inverse FFT at total cost
$O(N_\omega \log N_\omega)$, independent of $\lambda_j$.
The analytical gradient $(1+\partial_{\lambda_j})G/G$
\cref{eq:1plus_dlambda} is obtained at the cost of one additional
elementwise multiply by $T(\tilde{f}_\text{eff})$~\cref{eq:T_feff},
with no finite-difference overhead.
For a detector with $\mathcal{O}(10^4)$ channels and variable per-channel
$\lambda_j$, as in large-scale neutrino experiments, the pre-factorised
structure \cref{eq:G_factored} makes joint charge-time likelihood
evaluation \cref{eq:logL} tractable without storing or convolving
per-$n$ histograms, at a memory footprint independent of signal intensity.
The charge-model evaluation is expected to dominate per-channel likelihood
cost relative to rate-integral interpolation at high channel counts,
since the latter can be batched over a shared detector-response table.
This paper provides the likelihood term and its theoretical complexity;
full reconstruction performance---vertex resolution, energy bias, and
systematic uncertainties---requires a dedicated detector-level study
incorporating optical and transport models.

\section{Conclusion}\label{sec:conclusion}

We have derived a closed-form characteristic-function framework
for the SiPM charge response.
The three physical contributions (pedestal noise, signal charge,
and dark counts) factorise into independent multiplicative factors on a
single frequency grid, replacing the truncated triple sum
\cref{eq:direct_sum} with a closed-form $\mathcal{O}(N_\omega)$ expression
free of truncation parameters and combinatorial bookkeeping,
exact within the stated stochastic model.
Prompt internal optical cross-talk is handled via a Lambert~$W$
closed form for the Galton--Watson total-progeny PGF
\cref{eq:G_full_explicit}; the per-avalanche afterpulse chain is
derived as the maximum-entropy Poisson--Gamma mixture, whose
$N$-avalanche total is Negative Binomial; and dark counts enter as an
independent compound-Poisson factor \cref{eq:D_cf}, exact at all rates for the
stated independent compound-Poisson dark-count model.

Validation on Hamamatsu S13360-series PCB6 data across 53--60\,V
confirms BIC model selection resolves three identifiability regimes:
the baseline model (M1) suffices below 54\,V; DCR becomes identifiable
at 54\,V (M3); the full model (M4) is selected above 54.5\,V where
inter-PE valleys jointly constrain afterpulse and dark-count parameters.
The factorised $\lambda$-dependence \cref{eq:G_factored} yields a
pre-computable $h(\omega_k)$ array; applied to event-level reconstruction
(\cref{sec:reconstruction}), this reduces per-channel likelihood
evaluation to a single elementwise multiply and one inverse FFT,
with the analytical gradient \cref{eq:1plus_dlambda,eq:T_feff} at the
cost of one additional multiply---no finite-difference overhead and no
dependence on $k_\text{max}$ or signal intensity.

The validation covers the Hamamatsu S13360-series PCB6 device at room
temperature; parameters at other temperatures or with different SiPM
types require separate characterisation.
The minimal trigger efficiency $\eta(x)=x$ in the afterpulse charge model
can be replaced by the two-parameter form \cref{eq:ap_charge_general}
for detectors where charge-recovery and Geiger-efficiency recovery
timescales differ measurably.
Resolving individual afterpulse arrival times would further discriminate
trap species by release-time constants, and extending the framework to
temperature-dependent DCR and cross-talk parametrisations would
support cryogenic neutrino-detector deployment---both are left for future
work.

\section*{Acknowledgements}

The authors thank Hanwen Wang for insightful discussions on the statistical behaviors of silicon photomultipliers, and thank the JUNO Collaboration for its inspiration.
This work was supported by the National Key Research and
Development Program of China (Grant No.\ 2023YFA1606102, 2023YFC3107402).

\bibliographystyle{apsrev4-1}
\bibliography{references}

\end{document}